\documentclass[aps,prd,preprintnumbers,groupedaddress,nofootinbib,amssymb,notitlepage,eqsecnum]{revtex4-2}
\usepackage{here}
\usepackage[dvipdfmx]{graphicx}
\usepackage{amsmath,amsthm,amssymb}
\usepackage{bm}
\usepackage{color}
\usepackage{mathtools}

\usepackage{amsfonts}
\usepackage{dcolumn}
\usepackage{hyperref}
\allowdisplaybreaks[1]
\usepackage{stackengine}
\usepackage{ulem}

\usepackage{comment}

\newcommand{\be}{\begin{equation}}  
\newcommand{\ee}{\end{equation}}
\newcommand{\ba}{\begin{eqnarray}}
\newcommand{\ea}{\end{eqnarray}}

\newcommand{\rd}{{\rm d}}

\newcommand{\bem}{\begin{bmatrix}}
\newcommand{\eem}{\end{bmatrix}}
\newcommand{\Mpl}{M_{\rm Pl}}

\allowdisplaybreaks

\begin{document}

\preprint{YITP-23-106, WUCG-23-09}

\title{Starobinsky inflation with a quadratic Weyl tensor}

\author{Antonio De Felice$^{1,}$\footnote{{\tt antonio.defelice@yukawa.kyoto-u.ac.jp}}}
\author{Ryodai Kawaguchi$^{2,}$\footnote{{\tt ryodai0602@fuji.waseda.jp}}}
\author{Kotaro Mizui$^{2,}$\footnote{{\tt mizmiz8612@akane.waseda.jp}}}
\author{Shinji Tsujikawa$^{2,}$\footnote{{\tt tsujikawa@waseda.jp}}}

\affiliation{
$^1$Center for Gravitational Physics and Quantum Information, 
Yukawa Institute for Theoretical Physics, Kyoto University, 
606-8502 Kyoto, Japan\\
$^2$Department of Physics, Waseda University, 3-4-1 Okubo, Shinjuku, Tokyo 169-8555, Japan}

\begin{abstract}

In Starobinsky inflation with a Weyl squared Lagrangian 
$-\alpha C^2$, where $\alpha$ is a coupling constant, we study the linear stability of cosmological perturbations on a spatially flat Friedmann-Lema\^{i}tre-Robertson-Walker background. 
In this theory, there are two dynamical vector modes propagating as ghosts 
for $\alpha>0$. This condition is required to avoid tachyonic instabilities of vector perturbations during inflation. 
The tensor sector has four propagating degrees of freedom, among which two of them correspond to ghost modes. 
However, the tensor perturbations approach constants after the Hubble radius crossing during inflation, and hence the classical instabilities are absent. In the scalar sector, the Weyl curvature gives rise to a ghost mode coupled to the scalaron arising from the squared Ricci scalar. We show that two gauge-invariant gravitational potentials, which are both dynamical in our theory, are subject to exponential growth after the Hubble radius crossing. 
There are particular gauge-invariant combinations like the curvature perturbations whose growth is suppressed, but it is not possible to remove the instability of other propagating degrees of freedom present in the perturbed metric. 
This violent and purely classical instability present in the scalar sector makes the background unviable. Furthermore, the presence of such classical instability makes the quantization of the modes irrelevant, and the homogeneous inflationary background is spoiled by the Weyl curvature term.

\end{abstract}

\date{\today}


\maketitle

\section{Introduction}
\label{introsec}

The inflationary paradigm \cite{Starobinsky:1980te,Sato:1980yn,Kazanas:1980tx,Guth:1980zm} 
can successfully resolve several shortcomings in big bang cosmology, e.g., the horizon and flatness problems. Moreover, it can explain the origin of large-scale structures in the Universe by stretching quantum fluctuations over super-Hubble scales during the accelerated expansion \cite{Starobinsky:1979ty, Starobinsky:1982ee, Hawking:1982cz, Guth:1982ec, Bardeen:1983qw}. The spectra of scalar and tensor perturbations predicted in standard slow-roll inflation are consistent with the observed cosmic microwave background (CMB) temperature anisotropies. After the data release of WMAP \cite{WMAP:2003elm} and Planck \cite{Planck:2013pxb} satellites, we have been able to distinguish between many different models of inflation. In particular, the first model advocated by Starobinsky \cite{Starobinsky:1979ty} is still one of the best-fit models to the Planck CMB data \cite{Planck:2018jri}.

In the Starobinsky model, inflation is driven by a quadratic Ricci scalar term $\beta R^2$, where $\beta$ is a positive coupling constant. The period of cosmic acceleration ends when the $\beta R^2$ term drops below the Ricci scalar $R$ \cite{Starobinsky:1985ibc,Vilenkin:1985md,Mijic:1986iv,Tsujikawa:1999iv,Starobinsky:2001xq}.
The quadratic curvature scalar gives rise to a new propagating degree of freedom (d.o.f.) dubbed the ``scalaron'' \cite{Starobinsky:1979ty} with a mass squared $m_S^2=1/(6\beta)$ on the Minkowski background \cite{Olmo:2005hc, Faraoni:2005vk, Chiba:2006jp, Navarro:2005gh, Amendola:2007nt}.
Indeed, the $f(R)$ gravity given by the Lagrangian $f(R)=R+\beta R^2$ is equivalent to Brans-Dicke theory \cite{Brans:1961sx} with a scalaron potential arising from the gravitational sector \cite{OHanlon:1972xqa, Chiba:2003ir, Sotiriou:2008rp}.
From the observed amplitude of CMB temperature fluctuations, the scalaron mass is constrained to be $m_s \simeq 10^{-5}M_{\rm pl}$, where $M_{\rm pl}$ is the reduced Planck mass \cite{Mijic:1986iv, Mukhanov:1987pv, Salopek:1988qh, Hwang:2001pu}. 
The Starobinsky model predicts the scalar spectral index $n_s \simeq 1-2/N$ and the tensor-to-scalar ratio $r \simeq 12/N^2$ \cite{Mukhanov:1981xt, Hwang:2001pu, DeFelice:2010aj}, where $N$ is the number of $e$-foldings counted backward from the end of inflation. On scales relevant to the CMB observations ($N \simeq 55\sim 60$), the theoretical predictions of $n_s$ and $r$ are well-consistent with the Planck data combined with other data \cite{Planck:2018jri}.

From the viewpoint of an ultraviolet completion of gravity, there are also other quadratic curvature contributions to the Lagrangian constructed from 
scalar products of the Riemann tensor $R_{\mu \nu \rho \sigma}$ and the Ricci tensor $R_{\mu \nu}$ \cite{Salvio:2018crh}. 
Given that the Gauss-Bonnet curvature invariant ${\cal G}=R^2-4R_{\mu \nu}R^{\mu \nu}+R_{\mu \nu \rho \sigma}R^{\mu \nu \rho \sigma}$ is a topological term that does not affect the field equations of motion \cite{Lovelock:1971yv}, the general quadratic-order Lagrangian can be expressed in the form $L_2=-\alpha C^2+\beta R^2$, where $C^2=2R_{\mu \nu}R^{\mu \nu}-2R^2/3+{\cal G}$ is a squared Weyl curvature. 
This quadratic theory of gravity, which was originally advocated by Stelle \cite{Stelle:1976gc}, is renormalizable and also asymptotically free \cite{Fradkin:1981iu}. 
However, the Weyl curvature generally gives rise to ghost d.o.f.s associated with derivative terms higher than second order in the field equations of motion \cite{Stelle:1977ry}.

Albeit the appearance of ghosts in Weyl gravity with the Lagrangian $-\alpha C^2$, the perturbative expansion about the Minkowski vacuum shows that all the linear perturbations in scalar, vector, and tensor sectors propagate with the speed of light \cite{Hindawi:1995an, Bogdanos:2009tn, Hinterbichler:2015soa}.
This means that, in the absence of additional matter sources, the perturbations are not subject to Laplacian instabilities. 
Hence, on Minkowski, the Weyl ghosts can be of ``soft'' types \cite{Smilga:2004cy}, i.e.,\ even in the presence of the ghosts, the classical perturbations do not grow by either Laplacian or tachyonic instabilities on the given background. 
However, this situation should be different by introducing some matter fields or by taking into account the $\beta R^2$ term on curved backgrounds. The latter corresponds to Stelle's quadratic curvature theory mentioned above, in which case the scalaron field arising from the $\beta R^2$ term can be gravitationally coupled to the Weyl ghost.

In this paper, we would like to address the stability of linear perturbations on the inflationary background realized by Stelle's theory. We note that there are some related works in which the dynamics of cosmological perturbations during inflation were discussed in the presence of the Weyl curvature term \cite{Clunan:2009er, Deruelle:2010kf, Myung:2015vya, Ivanov:2016hcm, Anselmi:2021rye}.
Most of those papers assumed the existence of a canonical scalar field besides the Weyl and Einstein-Hilbert terms. Since the squared Weyl curvature is conformally invariant, the Lagrangian $L=R-\alpha C^2+\beta R^2$ of Stelle's theory can be transformed to that in the Einstein frame with kinetic and potential energies of the scalaron field as well as the Weyl term \cite{Whitt:1984pd, Barrow:1988xh, Maeda:1988ab, Gottlober:1989ww}.

The analysis of tensor perturbations on a spatially-flat Friedmann-Lema\^{i}tre-Robertson-Walker (FLRW) background \cite{Clunan:2009er} showed that the tensor ghosts can be soft during inflation in that they do not grow by 
classical instabilities. In the vector sector, there are two 
dynamical propagating d.o.f.s arising from the Weyl curvature term $-\alpha C^2$ \cite{Deruelle:2010kf}. 
The ghosts do not appear if $\alpha<0$, but vector perturbations are subject to tachyonic instabilities. For $\alpha>0$ the ghosts are present, but vector perturbations decay during inflation. 
Thus, despite the presence of ghosts for $\alpha>0$, both tensor and vector perturbations are not prone to classical Laplacian or tachyonic instabilities. 
These results were already recognized in 
Refs.~\cite{Clunan:2009er,Deruelle:2010kf} according to the analysis in 
the Einstein frame, but we will study whether a similar property holds 
in the Jordan frame. Indeed, for $\alpha>0$, the classical instabilities 
are absent for both the tensor and vector sectors.

In the scalar sector, the analysis of Ref.~\cite{Deruelle:2010kf} in the Einstein frame of Stelle's theory showed that gravitational potentials in a Newtonian gauge exhibit rapid growth after the Hubble radius crossing during inflation. 
On the other hand, the same paper also found that the curvature perturbation in a comoving gauge remains bounded.
In Ref.~\cite{Ivanov:2016hcm}, it was claimed that the former growth of gravitational potentials is a gauge artifact and that scalar perturbations are not prone to real instabilities. 
So far, it is not yet clear whether the instability in the scalar sector induced by the Weyl term corresponds to a real, physical one.
To clarify this issue, we need to scrutinize whether the instability of scalar perturbations generally persists or not independent of the gauge choices.

In this paper, we will study the evolution of cosmological perturbations during inflation 
in the Jordan frame of Stelle's theory by paying particular attention to the classical stability of the scalar modes. 
For this purpose, we choose several different physical gauges and analytically derive the closed fourth-order differential equations for the gravitational potentials as well as other gauge-invariant variables like curvature perturbations.
We will explicitly show that two dynamical propagating d.o.f.s arise from the $\beta R^2$ term (i.e., scalaron) and the Weyl curvature, one of which always behaves as a ghost mode. Therefore, in general, there are four independent initial conditions necessary to uniquely specify the classical evolution of the scalar sector, and that is the reason why the system can be described in terms of a closed fourth-order differential equation for one single scalar mode, or evidently, by two second-order differential equations for two independent fields.

We will show that the two gravitational potentials $\Psi$ and $\Phi$, which are both propagating d.o.f.s, exponentially grow after the Hubble radius crossing. This instability of $\Psi$ and $\Phi$ occurs independently of the gauge choice made to study their dynamics. 
Among other relevant gauge-invariant variables, we also find that the curvature perturbation is a specific variable approaching a constant in the large-scale limit. However, the exponential increase of at least one dynamical scalar d.o.f. appearing in the perturbed line element does not allow the FLRW spacetime to be a stable cosmological background. 
Thus, the inflationary background is violated by this real, physical, and classical instability of scalar perturbations induced by the Weyl ghost coupled to the scalaron. Therefore, we conclude the propagating ghost d.o.f. in the scalar sector is not 
of the soft type. Well before the end of inflation, the cosmological background 
is spoiled by the classical instability and it changes to a highly inhomogeneous 
Universe. In such a context, we believe that this lack of a homogeneous background makes the quantization of perturbations irrelevant. If the ghost modes were soft, then the quantization procedure would acquire relevance and the results could be interesting.
However, this is not the case for inflation in quadratic gravity with the Weyl term.

This paper is organized as follows.
In Sec.~\ref{backsec}, we briefly review the background dynamics of inflation realized in Stelle's theory.
In Sec.~\ref{vecsec}, we revisit how vector perturbations propagate as truly dynamical d.o.f.s 
and show that the absence of tachyonic instabilities requires the condition $\alpha>0$. 
In Sec.~\ref{tensec}, we see that, despite the appearance of ghosts 
arising from the Weyl term, the four dynamical d.o.f.s of 
tensor perturbations approach constants after the Hubble radius crossing.
In Sec.~\ref{scasec}, we study the evolution of scalar perturbations by choosing several different gauges and show that, independently of the gauge choices, two modes propagate and the FLRW background is spoiled by the presence of instabilities of at least one of the dynamical d.o.f.s present in the perturbed line element. 
Although the classical instability itself is present for any nonzero initial conditions of scalar modes, we confirm in Sec.~\ref{sec:numerics} its presence by numerically integrating the perturbation equations of motion with proper initial conditions.
Sec.~\ref{consec} is devoted to conclusions.

\section{Inflation in quadratic gravity}
\label{backsec}

The action in quadratic gravity contains scalar products of two contractions of the Riemann tensor $R_{\mu \nu \rho \sigma}$, Ricci tensor $R_{\mu \nu}$, and Ricci scalar $R$. On using the property that the Gauss-Bonnet term ${\cal G}=R^2-4R_{\mu \nu}R^{\mu \nu}+R_{\mu \nu \rho \sigma} R^{\mu \nu \rho \sigma}$ is topological, the Riemann products $R_{\mu \nu \rho \sigma}R^{\mu \nu \rho \sigma}$ can be eliminated from the action.
Taking the Einstein-Hilbert term $\Mpl^2 R/2$ into account, the action of quadratic gravity can be expressed in the form \cite{Stelle:1976gc}
\be
{\cal S}=\frac{M_{\rm pl}^2}{2}\int {\rm d}^4 x \sqrt{-g} 
\left( R -\alpha C^2 +\beta R^2 \right)\,,
\label{Saction}
\ee
where $g$ is a determinant of the metric tensor $g_{\mu \nu}$, $\alpha$ and $\beta$ are constants, and $C^2$ is the Weyl tensor squared given by 
\be
C^2=2R_{\mu \nu} R^{\mu \nu}-\frac{2}{3}R^2+{\cal G}\,.
\ee
Up to boundary terms, the action (\ref{Saction}) can be expressed as
\be
{\cal S}=\frac{M_{\rm pl}^2}{2}\int {\rm d}^4 x 
\sqrt{-g} 
\left[  R 
-2\alpha R_{\mu \nu} R^{\mu \nu}
+\left( \frac{2}{3}\alpha+\beta \right)R^2
\right] \,.
\label{Saction2}
\ee

We consider a spatially flat FLRW background described 
by the line element 
\be 
\rd s^2= -N^2(t) \rd t^{2}+a^2(t) \delta_{ij} \rd x^i \rd x^j\,,
\label{background}
\ee
where $a(t)$ is a time-dependent scale factor, and $N(t)$ is a lapse function. 
Varying the action (\ref{Saction2}) with respect to $N(t)$ and $a(t)$, respectively, and setting $N(t)=1$ at the end, it follows that 
\ba
& & 
H^2+6\beta \left( 6 H^2 \dot{H}-\dot{H}^2
+2H \ddot{H} \right)=0\,,\label{back1}\\
& & 
2\dot{H}+3H^2+6\beta \left( 18 H^2 \dot{H}
+9\dot{H}^2+12H \ddot{H}
+2\dddot{H} \right)=0\,,
\label{back2}
\ea
where $H=\dot{a}/a$ is the Hubble expansion rate, with a dot being the derivative with respect to $t$. 
At the background level, the Weyl curvature term does not contribute to the field equations of motion. 
This is an outcome of the conformal invariance of the Weyl curvature term, whose components vanish for the conformally flat background. 
Taking the time derivative of (\ref{back1}) and combining it with Eq.~(\ref{back1}), we obtain the same equation as (\ref{back2}).
This means that there is only a single independent equation of motion, Eq.~(\ref{back1}), governing the background dynamics.

During inflation, the Hubble expansion rate is nearly constant, and hence the last two terms in the parenthesis of Eq.~(\ref{back1}) are suppressed relative to the term $6 H^2 \dot{H}$. 
Then, there is the approximate relation 
\be
\dot{H} \simeq -\frac{1}{36 \beta} 
= -\frac{m_S^2}{6}\,,
\label{dotH}
\ee
where $m_S^2$ corresponds to the mass squared of a scalaron field given by \cite{Olmo:2005hc, Faraoni:2005vk, Chiba:2006jp, Navarro:2005gh, Amendola:2007nt}
\be
m_S^2=\frac{1}{6 \beta}\,.
\ee
Provided that 
\be
\beta>0\,,
\label{betacon}
\ee
there is no tachyonic instability arising from the negative value of $m_S^2$. Under the condition (\ref{betacon}), the Hubble parameter (\ref{dotH}) also decreases during inflation. 
We will impose the condition (\ref{betacon}) throughout the discussion below.

{}From Eq.~(\ref{dotH}), we obtain the following integrated solutions:
\ba
H(t) &\simeq& H_i-\frac{m_S^2}{6}(t-t_i)\,,
\label{Hso} \\
a(t) &\simeq& a_i \exp \left[ H_i (t-t_i) 
-\frac{m_S^2}{12}(t-t_i)^2 \right]\,,
\ea
where $H_i$ and $a_i$ are the values of $H$ and $a$ at the onset of inflation, respectively. 
We introduce the slow-roll parameter $\epsilon$, as
\be
\epsilon \equiv -\frac{\dot{H}}{H^2} 
\simeq \frac{m_S^2}{6H^2}\,.
\ee
The end of inflation is characterized by the Hubble parameter $H_f$ when $\epsilon$ becomes 
equivalent to 1, so that $H_f \simeq m_S/\sqrt{6}$. 
As we will see below, $H_i$ is larger than $m_S$. 
Then, by using Eq.~(\ref{Hso}), we can approximately estimate the time $t_f$ at the end of inflation, as $t_f \simeq t_i+6H_i/m_S^2$. 
The number of $e$-foldings acquired during inflation is given by 
\be
N \equiv \int_{t_i}^{t_f} H\,{\rm d}t
\simeq H_i (t_f-t_i)-\frac{m_S^2}{12}
(t_f-t_i)^2 \simeq \frac{3H_i^2}{m_S^2}\,.
\ee
Taking $N=60$ as a typical minimal $e$-folding number required to address the horizon and flatness problems, we obtain the initial Hubble parameter $H_i \simeq 4.5 m_S$.
This value translates to 
\be
\beta H_i^2 \simeq 3.4\,,
\ee
and hence $\beta H^2$ is of order 1 during inflation. 
From the viewpoint of ultraviolet completion of gravity, it is natural to consider the value of $|\alpha|$ same order as $\beta$. 

We note that quadratic gravity given by the action (\ref{Saction}) corresponds to 
the $f(R)=R+\beta R^2$ theory with the Weyl squared term $-\alpha C^2$. 
Under a conformal transformation of the metric tensor $g_{\mu \nu}$, the theory can be transformed to a metric frame described by Einstein's gravity in the presence of a scalaron field with the potential and the Weyl squared term  \cite{Whitt:1984pd, Barrow:1988xh, Maeda:1988ab, Gottlober:1989ww}.
In this Einstein frame, the dynamics of cosmological perturbations during inflation were carried out in Refs.~\cite{Clunan:2009er, Deruelle:2010kf, Myung:2015vya, Ivanov:2016hcm, Anselmi:2021rye} without necessarily relating the scalaron potential with the one arising from the original Lagrangian $f(R)=R+\beta R^2$. In this paper, we will perform all the analysis in the physical Jordan frame. 
To study the dynamics of perturbations during inflation, we do not need to take into account additional matter sources to the Jordan-frame action (\ref{Saction}).

Around the background (\ref{background}) with $N(t)=1$, we consider metric perturbations which depend on the cosmic time $t$ and spatial coordinates $x^i$. 
The perturbed line element is given by 
\be
\rd s^2=-\left(1+2A \right) \rd t^2
+2a(t) \left( \partial_i B+V_i \right) \rd t\,\rd x^i
+a^2 (t) \left[ (1+2\psi) \delta_{ij}
+2\partial_i \partial_j E
+\partial_i F_j+\partial_j F_i
+h_{ij} \right] \rd x^i \rd x^j\,,
\ee
where we used the notation $\partial_i=\partial/\partial x^i$, and the Latin indices represent spatial coordinates.
The four quantities $A$, $B$, $\psi$, and $E$ correspond to scalar perturbations, while $V_i$ and $F_i$ are vector perturbations satisfying the divergence-free conditions $\delta^{ij}\partial_j V_i=0$ and $\delta^{ij}\partial_j F_i=0$.
The intrinsic tensor perturbation is given by $h_{ij}$, which satisfies the traceless and transverse conditions ${h_i}^i=0$ and $\delta^{ik}\partial_k h_{ij}=0$.

\section{Vector perturbations}
\label{vecsec}

In quadratic gravity given by the action (\ref{Saction2}), we first study the dynamics of vector perturbations during inflation. Since there is the residual gauge d.o.f., we choose the following gauge condition 
\be
F_{i}=0\,,
\ee
where $i=1,2,3$. Then, the perturbed line element in the vector sector is given by 
\be
\rd s^2=-\rd t^2
+2a(t) V_i\,\rd t \rd x^i
+a^2 (t) \delta_{ij}
\rd x^i \rd x^j\,.
\ee
For practical computations, it is convenient to choose the vector-field configuration in the form 
\be
V_i=[V_1(t,z), V_2(t,z),0]\,,
\ee
which satisfies the divergence-free condition $\partial^i V_i=0$. 
Expanding the action (\ref{Saction2}) up to quadratic order in vector perturbations, integrating it by parts, and using the background Eq.~(\ref{back1}), the second-order action yields
\be
{\cal S}_v^{(2)}=-\frac{\Mpl^2}{2} \alpha \sum_{i=1,2}
\int \rd^4 x\,a
\left\{ \dot{U}_i^2-\frac{U_i'^2}{a^2}
-\left[ \frac{1}{2\alpha}
+\frac{6\beta}{\alpha}(2H^2+\dot{H}) 
\right]U_i^2 \right\}\,,
\label{Sv}
\ee
where a prime represents the derivative with respect to $z$, and $U_i \equiv V_i'$.
The Weyl curvature term gives rise to two dynamical vector perturbations $U_1$ and $U_2$.
For $\alpha>0$, it is clear from the action (\ref{Sv}) that both $U_1$ and $U_2$ behave as ghosts. 
On the other hand, the vector ghosts are absent if $\alpha<0$. 

Varying the action (\ref{Sv}) with respect to $U_i$ (with $i=1,2$), we obtain their equations of motion in real space. Then, we perform the Fourier transformation
\be
U_i=\frac{1}{(2\pi)^{1/2}} \int \rd k\,
\tilde{U}_i (t,k) 
e^{ik z}\,,
\ee
where $k$ is a comoving wave number, and $\tilde{U}_i$ is a function of $t$ and $k$. 
Omitting the tilde from $\tilde{U}_i(t,k)$ in the following, we obtain the vector perturbation equations of motion in Fourier space, as
\be
\ddot{U}_i+H \dot{U}_i+\left[ \frac{k^2}{a^2}
+m_{W}^2+\frac{6\beta}{\alpha}(2H^2+\dot{H}) 
\right]U_i=0\,,
\label{ddotU}
\ee
where 
\be
m_W^2 \equiv \frac{1}{2\alpha}\,,
\ee
is the mass squared arising from the Weyl curvature \cite{Stelle:1977ry, Hindawi:1995an, Bogdanos:2009tn, Hinterbichler:2015soa}. 
We note that the Lagrangian $\beta R^2$ contributes to the vector mass through the term $(6\beta/\alpha)(2H^2+\dot{H})$. 
During inflation ($|\dot{H}| \ll H^2$), the effective mass squared of vector perturbations is approximately given by 
\be
m_{\rm eff}^2 \simeq m_{W}^2+\frac{12\beta}{\alpha}H^2
=\frac{1+24\beta H^2}{2\alpha}\,.
\label{meff}
\ee
Since we are considering a positive coupling $\beta$, we have $m_{\rm eff}^2>0$ if $\alpha>0$ and $m_{\rm eff}^2<0$ if $\alpha<0$.

For the Weyl coupling constant in the range $|\alpha| \lesssim \beta$, the effective mass squared (\ref{meff}) is at least of order $H^2$.
Then, after $(k/a)^2$ drops below $|m_{\rm eff}^2|$ during inflation, Eq.~(\ref{ddotU}) approximately reduces to
\be
\ddot{U}_i+H \dot{U}_i+m_{\rm eff}^2\,U_i 
\simeq 0\,.
\ee
Since $H$ and $m_{\rm eff}^2$ can be approximated as constants during inflation, we obtain the following solution
\be
U_i=c_1 \exp \left( \frac{-H + \sqrt{H^2-4m_{\rm eff}^2}}{2}t \right)+c_2\exp \left( \frac{-H-\sqrt{H^2-4m_{\rm eff}^2}}{2}t \right)\,,
\label{U1}
\ee
where $c_1$ and $c_2$ are integration constants. 

For $\alpha>0$ (i.e., $m_{\rm eff}^2>0$), if $H$ is initially in the range $H>2m_{\rm eff}$, the amplitude of vector perturbations first decreases in proportion to $U_i \propto \exp[(-H+\sqrt{H^2-4m_{\rm eff}^2})t/2]$.
After $H$ decreases below $2m_{\rm eff}$, the vector perturbation starts to oscillate with the decreasing amplitude ($|U_i| \propto e^{-Ht/2}$). 
This means that, even though the two vector ghosts are present for $\alpha>0$, vector perturbations decay exponentially during inflation.

For $\alpha<0$, the negative mass squared $m_{\rm eff}^2$ leads to the tachyonic growth of $U_i$. 
During inflation, the first term on the right-hand side of Eq.~(\ref{U1}) corresponds to a growing mode. 
The increase of $U_i$ is particularly prominent after $H^2$ drops below the order of $-m_{\rm eff}^2$, during which $U_i \propto \exp(\sqrt{|m_{\rm eff}^2|}\,t)$. 
Then, for $\alpha<0$, the FLRW background is destroyed by the growth of vector perturbations.
To avoid such an instability problem, we will focus on the coupling in the range 
\be
\alpha>0\,,
\ee
in the following.

We recall that, in the above discussion, we have focused on the case $|\alpha| \lesssim \beta$. 
In the coupling range $|\alpha| \gg \beta$ with $\beta H_i^2$ of order 1, $|m_{\rm eff}^2|$ is much smaller than $H^2$ during inflation. 
Taking the limit $|m_{\rm eff}^2|/H^2 \to 0$ in Eq.~(\ref{U1}), the rapid growth of $U_i$ is absent.
In Sec.~\ref{scasec}, however, we will show that scalar perturbations are subject to exponential instabilities in the coupling range $|\alpha| \gg \beta$. 

\section{Tensor perturbations}
\label{tensec}

We proceed to study the dynamics of 
tensor perturbations $h_{ij}$ with the perturbed line element given by 
\be
\rd s^2=-\rd t^2+a^2(t) \left( \delta_{ij}+h_{ij} 
\right) \rd x^i \rd x^j\,,
\ee
with the traceless and transverse conditions 
${h_i}^i=0$ and $\partial^i h_{ij}=0$. 
Without loss of generality, we can consider 
gravitational waves propagating along the $z$ 
direction, whose nonvanishing components are 
\begin{equation}
h_{11} = -h_{22} = \frac{h_1(t,z)}{\sqrt{2}}\,,
\qquad
h_{12} = h_{21} = \frac{h_2(t,z)}{\sqrt{2}}\,,
\end{equation}
where $h_1$ and $h_2$ are functions of $t$ and $z$. 
These components of $h_{ij}$ automatically satisfy
the traceless and transverse conditions mentioned above.

Expanding the action (\ref{Saction2}) up to second order in $h_1$ and $h_2$ and integrating the action by parts, we obtain the following quadratic-order action 
\be
{\cal S}_t^{(2)}=\frac{\Mpl^2}{4}\int {\rm d}^4 x\, a^3 \sum_{i=1,2} 
\left[ -\alpha \ddot{h}_i^2-\frac{\alpha}{a^4} h_i''^2
+\frac{2\alpha}{a^2} \dot{h}_i'^2+\left\{ \frac{1}{2} 
+\left(\alpha+6 \beta \right) \left( 2H^2+\dot{H} \right) \right\} 
\dot{h}_i^2- \left\{ \frac{1}{2}
+6 \beta \left( 2H^2+\dot{H} \right) \right\} 
\frac{h_i'^2}{a^2} 
\right]\,, 
\label{St}
\ee
where we also used the background Eq.~(\ref{back1}).
The presence of the Weyl curvature term  
gives rise to the fourth-order differential equation for $h_i$. We perform the Fourier transformation of $h_i$, as 
\be
h_i=\frac{1}{(2\pi)^{1/2}} \int \rd k\,\tilde{h}_i (t,k) 
e^{ik z}\,,
\ee
where $\tilde{h}_i$ depends on $t$ and the wave number $k$.
In Fourier space, the second-order action 
$\tilde{{\cal S}}_t^{(2)}=\int \rd t \rd^3k \,L_{t}$ 
can be obtained under the replacements $\ddot{h}_i^2 \to \ddot{\tilde{h}}_i^2$, 
$h_i''^2 \to k^4 \tilde{h}_i^2$, $\dot{h}_i'^2 \to k^2 \dot{\tilde{h}}_i^2$, $\dot{h}_i^2 \to \dot{\tilde{h}}_i^2$, and 
$h_i'^2 \to k^2 h_i^2$ in Eq.~(\ref{St}).
Omitting the tilde from $\tilde{h}_i$, the second-order 
Lagrangian density in Fourier space is given by 
\be
L_t=\frac{a^3 \Mpl^2}{4} \sum_{i=1,2} 
\left[ -\alpha \ddot{h}_i^2+\left\{ \frac{1}{2} 
+\left(\alpha+6 \beta \right) \left( 2H^2+\dot{H} \right)
+\frac{2\alpha k^2}{a^2} \right\} 
\dot{h}_i^2- \frac{k^2}{a^2} \left\{ \frac{1}{2}
+6 \beta \left( 2H^2+\dot{H} \right)+\frac{\alpha k^2}{a^2} \right\}
h_i^2 \right]. 
\label{Lt}
\ee

To understand the appearance of ghost d.o.f.s, we introduce Lagrangian multipliers $\chi_i$ (with $i=1,2$) such that 
\begin{equation}
\bar{L}_{t}=L_{t} +\frac{a^3 \Mpl^2 \alpha }{4} \sum_{i=1,2}
\left( \ddot{h}_i+c_{1} {\dot h}_i+c_{2} h_i-c_{3} \chi_i 
\right)^{2}\,,
\label{bLt}
\end{equation}
where $c_i$'s are time-dependent coefficients. 
We note that the coefficient $a^3 \Mpl^2 \alpha/4$ 
in front of $\ddot{h}_i^2$ has been introduced to cancel 
the first term in Eq.~(\ref{Lt}). 
We fix $c_i$'s in Eq.~(\ref{bLt}) 
to obtain the Lagrangian density containing a canonical 
form of the kinetic terms $\dot{h}_i$ and $\dot{\chi}_i$, 
without the product $\dot{h}_i \chi_i$.
For this purpose, we choose
\begin{align}
c_1= 3H\,,\qquad
c_2 = \frac{k^{2}}{a^{2}}
+\frac{4\alpha (H^2-\dot{H})+12\beta (2H^2+\dot{H})-3}{4\alpha}\,,\qquad
c_3 =\frac{2}{\alpha} \,.
\label{c123}
\end{align}
After integration by parts, the Lagrangian density (\ref{bLt}) reduces to
\ba
\bar{L}_t &=& \frac{1}{2}\,a^3 \Mpl^2 \sum_{i=1,2} 
\Biggl( {\dot h}_i^2+2{\dot h}_i {\dot\chi}_i+
\biggl[ 3\beta (H^2 - \dot{H})(2 H^2 + \dot{H})
-\frac{5H^2}{4}+\frac{\dot{H}}{2}-\frac{k^2}{a^2}
+\frac{9\{ 1 - 4 \beta (2 H^2 +\dot{H})\}^2}{32\alpha} 
\nonumber \\
& &\qquad \qquad \qquad 
+\frac{\alpha}{12} \left\{ 6H^2 (H^2-2\dot{H})
-\frac{1}{\beta}(H^2-\dot{H})-12 \dot{H}
\left( \frac{2k^2}{a^2}-5\dot{H} \right) \right\}\biggr] 
h_{i}^{2} 
+\frac{2}{\alpha} \chi_i^2
\nonumber \\
& &\qquad \qquad \qquad 
-\left\{ 2 H^2-2\dot{H} +\frac{2k^2}{a^2}
+\frac{3[4\beta(2H^2+\dot{H})-1]}{2\alpha} \right\} 
h_i \chi_i
\Biggr)\,.
\label{Ltr}
\ea
{}From this expression, we find that there are four dynamical 
perturbations $h_1$, $h_2$, $\chi_1$, and $\chi_2$ in the tensor sector, 
in agreement with the analysis of Refs.~\cite{Clunan:2009er,Bogdanos:2009tn,Deruelle:2010kf}.
Terms containing the product of time derivatives in Eq.~(\ref{Ltr}) 
can be expressed in the form 
\be
(\bar{L}_t)_K=\sum_{i=1,2} \dot{\vec{\psi}}_i {\bm K} \dot{\vec{\psi}}_i^{\,{\rm T}}\,,
\ee
where $\vec{\psi}_i=(h_i, \chi_i)$, and ${\bm K}$ is a 
$2 \times 2$ symmetric matrix whose components are 
\be
K_{11}=\frac{1}{2}\,a^3 \Mpl^2\,,\qquad 
K_{12}=K_{21}=\frac{1}{2}\,a^3 \Mpl^2\,,\qquad 
K_{22}=0\,.
\ee
The absence of ghosts requires the following conditions
\be
K_{11}=\frac{1}{2}\,a^3 \Mpl^2>0\,,\qquad {\rm and} 
\qquad 
-K_{12}^2=-\frac{1}{4}\,a^6 \Mpl^4>0\,.
\ee
While the former is satisfied, the latter is always 
violated. Hence the two ghosts are present, 
besides the other two no-ghost modes.

In the following, we will study the propagation 
of tensor perturbations during inflation.
Varying the Lagrangian density (\ref{bLt}) with respect 
to $\chi_i$, it follows that: 
\begin{equation}
\chi_i=\frac{1}{2} \alpha \ddot{h}_i  
+\frac{3}{2}\alpha H \dot{h}_i 
+\frac12\left[\frac{\alpha k^{2}}{a^{2}}
+\alpha (H^2 -\dot{H})+3\beta (2H^2 +\dot{H})
-\frac{3}{4}
\right] h_{i}\,,
\label{chii}
\end{equation}
where we used the coefficients (\ref{c123}). 
We also vary Eq.~(\ref{Ltr}) with respect to $h_i$ 
and eliminate $\chi_i$ and their time derivatives by exploiting 
Eq.~(\ref{chii}). 
Then, we obtain the following fourth-order differential equation
\begin{align}
\ddddot{h_i}&=
-6 H \dddot{h_i}-\left[\frac{2 k^{2}}{a^{2}}
+\frac{\left( 4 \alpha+6 \beta \right) {\dot H}}{\alpha}
+\frac{1+\left( 22 \alpha+24 \beta \right) H^{2}}
{2 \alpha}\right] \ddot{h_i} \nonumber\\
&\quad-\left[\frac{2 H k^{2}}{a^{2}}
+\frac{\left(\alpha+6\beta \right) {\dot H}^2}{2\alpha  H}
+\frac{4(\alpha +6\beta) H {\dot H}}{\alpha}
+\frac{H \{ 72\beta (\alpha+6 \beta)H^2 -\alpha +12 \beta \}}
{12 \alpha  \beta}\right] \dot{h_i} \nonumber\\
&\quad-\left[\frac{k^{4}}{a^{4}}
+\left( \frac{1+24 \beta H^{2}}{2\alpha}
+\frac{6 \beta  {\dot H}}{\alpha}\right) \frac{k^{2}}{a^2}
\right] h_{i}\,.\label{eq:dh}
\end{align}
We note that the same equation also follows by directly varying the 
original Lagrangian density (\ref{Lt}) with respect to $h_i$.
Equation (\ref{eq:dh}) governs the dynamics of tensor perturbations. 

We first solve Eq.~(\ref{eq:dh}) in the high-momentum regime, namely,
for the modes deep inside the Hubble radius ($k/a\gg H$). 
After the Hubble radius crossing during inflation, the perturbations 
enter the super-Hubble region $k/a<H$.
The evolution of $h_i$ in the latter 
large-scale regime will be discussed later.
Keeping only the most dominant terms for sub-Hubble perturbations 
and expressing the Fourier components as 
$h_i(t,k)=\tilde{h}_i (t) 
e^{-i \int \omega (t,k) \rd t}$, 
where $\tilde{h}_i$ is a function of $t$, and $\omega$ depends on 
$t$ and $k$, the Wentzel-Kramers-Brillouin (WKB) approximation gives the relation 
$\dot{h}_i \simeq -i\omega h_i$.
In this WKB regime, we also have the inequality 
$|\dot\omega|\ll \omega^2$. 
Then, Eq.~\eqref{eq:dh} approximately reduces to
\begin{equation}
\omega^{4}+6i H \omega^{3}-\frac{2 k^{2} \omega^{2}}{a^{2}}
-\frac{2 i H k^{2} \omega}{a^{2}}+\frac{k^{4}}{a^{4}}\simeq0\,.
\label{dispe}
\end{equation}
We search for solutions of the kind $\omega=c_t\,k/a$, 
where $c_t$ is the tensor propagation speed. 
Substituting this dispersion relation into Eq.~(\ref{dispe}) 
and taking the small-scale limit $k/(aH) \gg 1$, 
it follows that 
\begin{equation}
\left( c_t^2-1 \right)^2\,\frac{k^4}{a^4}\,
\left[ 1+\mathcal{O} \left( \frac{aH}{k} 
\right) \right] = 0\,.
\end{equation}
At leading order in the expansion of the small 
parameter $aH/k$, we obtain
\be
c_t^2=1\,,
\ee
for all the four dynamical modes $h_1$, $h_2$, $\chi_1$, and $\chi_2$. 
Since the propagation speeds are luminal, there are no classical 
Laplacian instabilities in the tensor sector for perturbations 
deep inside the Hubble radius.

Let us consider the evolution of super-Hubble tensor modes
after the Hubble radius crossing, i.e., $k/a \ll H$. 
Since $|\dot{H}| \ll H^2$ during slow-roll inflation, 
Eq.~(\ref{eq:dh}) approximately reduces to
\begin{equation}
\ddddot{h_i}+6 H \dddot{h_i}
 +12\lambda_1 H^2 \ddot{h_i}
 +36\lambda_2 H^3 \dot{h_i}\simeq0\,,
\label{eq:GWs}
\end{equation}
where
\be
\lambda_1=\frac{(22 \alpha+24 \beta) H^{2}+1}
{24 \alpha H^2}\,,\qquad 
\lambda_2=\frac{72\beta (\alpha+6 \beta)H^2 -\alpha +12 \beta}
{432 \alpha  \beta H^2}\,.
\ee
The difference between $\lambda_1$ and $\lambda_2$ 
is given by  
\be
\lambda_2-\lambda_1=-\frac{3}{4}-\frac{1}{72\alpha H^2}
-\frac{1}{432 \beta H^2}\,.
\label{lam12}
\ee
For the coupling constants $\alpha$ and $\beta$ with 
$\alpha H^2 \gtrsim {\cal O}(1)$ and $\beta H^2 \gtrsim {\cal O}(1)$ during inflation, we can neglect the last two terms in Eq.~(\ref{lam12}) relative to $-3/4$. On using the approximate relation 
$\lambda_2 \simeq \lambda_1-3/4$ in this case and assuming 
that $H$ is constant during inflation, the solution 
to Eq.~(\ref{eq:GWs}) can be expressed as
\begin{equation}
h_i=A_{i}+B_{i}\,e^{-3Ht}+C_i\, e^{-[3-\sqrt{45-48\lambda_1}]Ht/2}
+D_i\, e^{-[3+\sqrt{45-48\lambda_1}] Ht/2}\,,
\label{hiso}
\end{equation}
where $A_i$, $B_i$, $C_i$, and $D_i$ are 
integration constants, and
\be
\lambda_1 \simeq \frac{11}{12}+\frac{\beta}{\alpha}
>\frac{11}{12}\,.
\ee
In the last inequality, we exploited the fact that both $\alpha$ 
and $\beta$ are positive.\footnote{In the limit $\alpha\gg\beta$ and $\beta H^2 \gtrsim \mathcal{O}(1)$, we find that $h_i=A_i+B_i e^{-3Ht}+C_ie^{-Ht}+ D_i e^{-2Ht}$, so that $h_i$ tends to $A_i$ also in this case.} 
For $11/12<\lambda_1\le 15/16$, the last three terms in 
Eq.~(\ref{hiso}) decrease exponentially.  
For $\lambda_1>15/16$, the last two terms in Eq.~(\ref{hiso}) exhibit damped oscillations with a decreasing amplitude proportional to $e^{-3Ht/2}$. 
This means that $h_i$ approaches the constant value $A_i$. 

If we consider the small Weyl coupling constant 
$\alpha \ll \beta$ with $\beta H^2 \gtrsim {\cal O}(1)$, 
then we have $\lambda_1 \simeq \lambda_2 \simeq 
\beta/\alpha \gg 1$. In the limit that 
$\beta/\alpha \to \infty$, the solution to 
Eq.~(\ref{eq:GWs}) is given by 
\be
h_i=A_{i}+B_{i}\,e^{-3Ht}+C_i\, 
e^{-( 3+i \Omega )Ht/2}
+D_i\, e^{-( 3-i \Omega )Ht/2}\,,
\label{hiso2}
\ee
where $\Omega$ is a constant. In this case, the amplitude of $h_i$ 
decreases in time as well and finally reaches a constant, $A_i$.

We have thus shown that, despite the appearance of ghosts, the tensor perturbation 
does not exhibit rapid growth during inflation.
In other words, the higher-order derivatives of $h_i$ appearing in the action (\ref{St}) hardly affect the standard conservation property of $h_i$, after the Hubble radius crossing.

\section{Scalar perturbations}
\label{scasec}

Let us next study the evolution of scalar perturbations for the perturbed line element given by  
\be
\rd s^2=-\left(1+2A \right) \rd t^2
+2a(t) \partial_i B  \rd t \rd x^i
+a^2 (t) \left[ (1+2\psi) \delta_{ij}
+2\partial_i \partial_j E \right] 
\rd x^i \rd x^j\,.
\ee
We consider an infinitesimal-gauge transformation 
\be
\tilde{t}=t+\xi^0\,,\qquad 
\tilde{x}^i=x^i+\delta^{ij} \partial_j \xi\,,
\label{gatra}
\ee
from one coordinate $x^{\mu}=(t,x^i)$ to another coordinate $\tilde{x}^{\mu}=(\tilde{t},\tilde{x}^i)$, where $\xi^0$ and $\xi$ are scalar quantities. 
Then, the four scalar perturbations $A$, $B$, $\psi$, and $E$ transform, respectively, as \cite{Bardeen:1980kt, Kodama:1984ziu, Mukhanov:1990me, Bassett:2005xm}
\be
\tilde{A}=A-\dot{\xi}^0\,,\qquad
\tilde{B}=B+\frac{\xi^0}{a}-a \dot{\xi}\,,\qquad
\tilde{\psi}=\psi-H \xi^0\,,\qquad
\tilde{E}=E-\xi\,.
\ee
The gauge-invariant gravitational potentials are defined by \cite{Bardeen:1980kt} 
\be
\Psi=A+\frac{\rd}{\rd t} \left[ 
a(B -a\dot{E}) \right]\,,\qquad 
\Phi=\psi+aH \left( B-a\dot{E} \right)\,.
\label{gravi}
\ee
We can also construct the following gauge-invariant variables
\be
\mathcal{A}\equiv A-\frac{\rd}{\rd t}\!
\left( \frac{\psi}{H} \right) \,,\qquad
\mathcal{B}\equiv aB+\frac\psi{H}-a^2\dot E\,.
\label{eq:FG}
\ee
While ${\cal B}$ is related to $\Phi$ as ${\cal B}=\Phi/H$, $\mathcal{A}=\Psi-(\rd/{\rd t})\!(\Phi/H)$ is not 
proportional to $\Psi$.

It is known that $f(R)$ gravity with nonlinear functions of $R$ gives rise to a scalar d.o.f. $\phi=\rd f/\rd R$ \cite{Starobinsky:1979ty, DeFelice:2010aj}.  
The quadratic action (\ref{Saction}) contains the function $f(R)=R+\beta R^2$, in which case $\phi=1+2\beta R$. 
Then, the perturbation of the new scalar d.o.f. is equivalent to $\delta \phi=2\beta \delta R$. 
We can construct a gauge-invariant quantity $\zeta=\psi-H \delta \phi/\dot{\phi}$ \cite{Lukash:1980iv, Lyth:1984gv}, or, equivalently,\footnote{Note that the field $\zeta$ is not well-defined on 
an exact de Sitter space where $H$ is constant. In the background for this model, the Hubble expansion rate varies due to the slow-roll 
evolution of the scalaron d.o.f., so we can still introduce $\zeta$.}
\be
\zeta=\psi-\frac{H}{\dot{R}} \delta R\,.
\label{zeta}
\ee
There is also the following combination analogous to the Mukhanov-Sasaki variable \cite{Mukhanov:1985rz, Sasaki:1986hm}:
\be
\delta R_f=\delta R-\frac{\dot{R}}{H}\psi\,.
\ee
which is related to $\zeta$, as $\delta R_f=-\dot{R}\zeta/H$.

We expand the action (\ref{Saction2}) up to quadratic order in scalar perturbations without fixing gauges and then derive the field equations of motion by varying the second-order action with respect to $A$, $B$, $\psi$, and $E$. 
These perturbation equations of motion are written in a gauge-ready form \cite{Hwang:2001qk, Heisenberg:2018wye, Kase:2018aps}, in that they are ready for the reader to choose a particular gauge. 
To fix the spatial part of the gauge transformation vector $\xi^{\mu}$, 
we choose the gauge 
\be
E=0\,.
\ee

In the following, we will work in Fourier space with the three 
dimensional comoving wave number ${\bm k}$. 
We omit a tilde for perturbed quantities 
in the Fourier space. 
Then, the perturbation of the Ricci scalar is given by 
\be
\delta R=
6 \left(\ddot{\psi}+4H \dot{\psi} \right)
+\frac{4k^2}{a^2} \psi
-6H \dot{A}-12 (2H^2+\dot{H})A
+\frac{2k^2}{a^2} A
+\frac{2}{a} k^2 \left( \dot{B}
+3H B \right)\,,
\label{deltaR}
\ee
where $k=|{\bm k}|$.
For the temporal part of $\xi^{\mu}$, we can consider several different gauge choices, including (A) Newtonian gauge ($B=0$), (B) flat gauge ($\psi=0$), and (C) unitary gauge ($\delta R=0$).    
The physical results, including the stability conditions and the evolution of scalar perturbations, are independent of the gauge choices. 

We first study conditions for the absence of ghosts and then proceed to address the dynamics of gauge-invariant perturbations for several different gauge choices.

\subsection{No-ghost conditions}
\label{ngsec}

In the flat gauge with $\psi=0$ and $E=0$, the gauge-invariant variables in Eq.~(\ref{eq:FG}) reduce to ${\cal A}=A$ and ${\cal B}=aB$, respectively.
We expand the action (\ref{Saction2}) up to quadratic order in two perturbations ${\cal A}$ and ${\cal B}$. 
In Fourier space, the second-order perturbed scalar Lagrangian density $L_s$ contains products of the time derivatives of ${\cal A}$ and ${\cal B}$ in the form 
\be
L_s \supset
K_{11}\dot{\cal A}^2
+K_{22} \dot{\cal B}^2
+2K_{12}\dot{\cal A}\dot{\cal B}\,,
\label{LsK}
\ee
with the coefficients
\be
K_{11}=18a^3\beta H^2\Mpl^2\,,\qquad
K_{22}=-\frac{2 (\alpha-3\beta)\Mpl^2 k^4}{3a}\,,\qquad 
K_{12}=-6a \beta H \Mpl^2 k^2\,.
\ee
Thus, both the fields ${\cal A}$ and ${\cal B}$ propagate as dynamical perturbations. 
The absence of ghosts requires the following two conditions 
\ba
K_{11} &=& 
18a^3\beta H^2\Mpl^2>0\,,
\label{NG1}\\
K_{11}K_{22}-K_{12}^2
&=&-12a^2 \alpha \beta 
H^2 \Mpl^4 k^4>0\,.
\label{NG2}
\ea
The kinetic term $K_{11}\dot{\cal A}^2$ corresponds to that of the scalaron perturbation $\mathcal{A}$, which does not behave as a ghost for $\beta>0$.
Under the no-ghost condition $\beta>0$ of the scalaron, the second equality (\ref{NG2}) is always violated for $\alpha>0$. 
Hence the other propagating DOF corresponds to a ghost mode, which is induced by the presence of the Weyl curvature term. 
To see this property more explicitly, we define the following fields that make the kinetic matrix diagonal 
\be
{\cal A}_1 \equiv {\cal A}+\frac{K_{12}}{K_{11}}{\cal B}\,,
\qquad 
{\cal B}_1 \equiv {\cal B}\,.
\label{fre1}
\ee
Then, the products of the time derivative of the new fields can be expressed in the form 
\be
L_s \supset
K_{11}\dot{\cal A}_1^2+\frac{K_{11}K_{22}-K_{12}^2}
{K_{11}} \dot{\cal B}_1^2
=a^3 \left( 18\beta H^2\Mpl^2 \dot{\cal A}_1^2
-\frac{2\Mpl^2 \alpha k^4}{3a^4} \dot{\cal B}_1^2
\right)\,.
\label{LsK2}
\ee
For positive values of $\alpha$ and $\beta$, we also introduce the canonically normalized fields 
\be
{\cal A}_2 \equiv 6\Mpl H \sqrt{\beta} {\cal A}_1\,,
\qquad 
{\cal B}_2 \equiv
\Mpl \sqrt{\frac{4\alpha}{3}} 
\frac{k^2}{a^2}{\cal B}_1\,,
\label{B2def}
\ee
so that the kinetic products of the Lagrangian density can be expressed as 
\be
(L_s)_K=a^3 \left( \frac{1}{2}\dot{{\cal A}}_2^2
-\frac{1}{2}\dot{{\cal B}}_2^2 \right)\,. 
\label{cano}
\ee
From this expression, it is clear that ${\cal A}_2$ and ${\cal B}_2$ are the canonically normalized perturbations corresponding to the no-ghost 
scalaron field and the Weyl scalar ghost, respectively.

In the above discussion we have chosen the flat gauge, but independent of the gauge choices, the scalar sector contains one ghost and the other no-ghost d.o.f.s.
In summary, for positive values of $\alpha$ and $\beta$, there are two vector ghosts, two tensor ghosts, and one scalar ghost among the total eight propagating d.o.f.s.\footnote{The recent analysis of black hole perturbations in Weyl gravity without the $\beta R^2$ term \cite{DeFelice:2023kpl} shows that there are seven dynamical d.o.f.s on a static and spherically symmetric background. Adding the $\beta R^2$ term gives rise to one scalar d.o.f., so the total dynamical d.o.f.s match each other on two different backgrounds.}
In the following, we study the evolution of scalar perturbations in detail by choosing three different gauges.

\subsection{Newtonian gauge}\label{sec:newton}

The Newtonian gauge corresponds to setting $B=0$ and $E=0$.
In this case the gauge-invariant gravitational potentials in Eq.~(\ref{gravi}) reduce to $\Psi=A$ and $\Phi=\psi$, so the perturbed line element is given by 
\be
{\rm d}s^2=-\left( 1+2\Psi \right) {\rm d}t^2
+a^2(t) \left( 1+2\Phi \right) \delta_{ij}
{\rm d}x^i {\rm d}x^j\,.
\label{Newtonian}
\ee
In the Newtonian gauge, the coordinate transformation vector $\xi^{\mu}$ is fixed on the FLRW background without any singularity. 
For this gauge choice, both $\Psi$ and $\Phi$ play the role of two dynamical perturbations.
{}From Eqs.~(\ref{zeta}) and (\ref{deltaR}), the relation between $\zeta$ and the gravitational potentials is 
\be
\zeta=\Phi-\frac{H}{\dot{R}} \left[ 
6 (\ddot{\Phi}+4H \dot{\Phi})
+\frac{4k^2}{a^2} \Phi-6H \dot{\Psi}
-12(2H^2+\dot{H})\Psi
+\frac{2k^2}{a^2} \Psi \right]\,.
\label{zeta2}
\ee

Varying the second-order action of scalar perturbations with respect to $A$, $B$, $\psi$, and $E$, we obtain the four perturbation equations of motion, respectively, for which we express in the form 
\be
{\cal E}_{A}=0\,,\qquad
{\cal E}_{B}=0\,,\qquad
{\cal E}_{\psi}=0\,,\qquad
{\cal E}_{E}=0\,,
\label{pereq}
\ee
After setting $B=0=E$ in the end, two of the above equations are independent, but the other two equations can be also used to obtain the closed differential equations for $\Psi$ and $\Phi$.

For example, the fourth-order differential equation for $\Psi$ can be derived by the following procedure.
We first solve the two equations ${\cal E}_{A}=0$ and ${\cal E}_{B}=0$ for $\dddot{\Phi}$ and $\ddot{\Phi}$.
Taking the time derivative of the latter and combining it with the former, we can eliminate the term $\dddot{\Phi}$ to obtain the other equation containing $\ddot{\Phi}$. Then, we can solve for $\dot{\Phi}$ and $\ddot{\Phi}$ by combining the two equations containing $\ddot{\Phi}$.
Performing a similar procedure further, we can express $\dot{\Phi}$ and $\Phi$ in terms of the derivatives of $\Psi$ up to third order. Taking the time derivative of the $\Phi$ equation and eliminating the $\dot{\Phi}$ term, we obtain the fourth-order differential equation for $\Psi$ in the form 
\be
\ddddot{\Psi}+\mu_1 H \dddot{\Psi}+\mu_2 H^2 
\ddot{\Psi}+\mu_3 H^3 \dot{\Psi}
+\mu_4 H^4\Psi=0\,,
\label{Psieq}
\ee
where $\mu_{1,2,3,4}$ are time-dependent dimensionless functions 
containing the $k$ dependence.
Due to its complexities, we do not write the explicit forms of these coefficients.

Similarly, we can derive the fourth-order differential equation 
for $\Phi$ in the form 
\ba
\ddddot{\Phi}+\nu_1 H \dddot{\Phi}+\nu_2 H^2 
\ddot{\Phi}+\nu_3 H^3 \dot{\Phi}
+\nu_4 H^4\Phi=0\,,
\label{Phieq}
\ea
where the functions $\nu_i$ ($i=1,2,3,4$) are not the same as $\mu_i$, respectively.

For the modes deep inside the Hubble radius ($k \gg aH$), 
the coefficients in Eqs.~(\ref{Psieq}) and (\ref{Phieq}) 
reduce to 
\be
\mu_1 \simeq \nu_1 \simeq 6\,,\qquad 
\mu_2 \simeq \nu_2 \simeq 
\frac{2k^2}{(aH)^2}\,,
\qquad 
\mu_3 \simeq \nu_3 \simeq 
\frac{2k^2}{(aH)^2}\,, \qquad
\mu_4 \simeq \nu_4 \simeq 
\frac{k^4}{(aH)^4}\,.
\label{mui}
\ee
In the WKB regime, we substitute the solutions 
$\Psi=\Psi_0 e^{-i \int \omega \rd t}$ and 
$\Phi=\Phi_0 e^{-i \int \omega \rd t}$
into Eqs.~(\ref{Psieq}) and (\ref{Phieq}) 
with the coefficients (\ref{mui}), where 
$\Psi_0$ and $\Phi_0$ are constants. 
This process leads to the same relation 
as Eq.~(\ref{dispe}). 
Writing the dispersion relation as $\omega=c_s k/a$ 
and taking the limit $k/a \gg H$, we obtain the 
squared propagation speeds  
\be
c_s^2=1\,,
\ee
for both $\Psi$ and $\Phi$.
This means that, for the modes deep inside the Hubble 
radius, the Laplacian instabilities are absent for 
the two gravitational potentials.

For super-Hubble modes, we take the limit $k/(aH) \ll 1$ 
in the coefficients $\mu_i$ and $\nu_i$. 
Moreover, we also take the limit where the slow-roll parameter 
$\epsilon=-\dot{H}/H^2$ goes to 0. 
Then, the coefficients are simplified to
\ba
\mu_1 &\simeq& \frac{864\beta^2 (\alpha + \beta) H^4
+18 \beta (\alpha + 4 \beta) H^2-\alpha + 3\beta}
{6 \beta H^2[48 \beta (\alpha + 3\beta) H^2+\alpha 
+ 12\beta]}\,,
\label{mu1}\\
\mu_2 &\simeq& -\frac{288 (\alpha - 12 \beta) \beta^2 
(\alpha + 3 \beta) H^4- 6 \beta 
(7\alpha^2 + 36\alpha \beta 
+ 216 \beta^2) H^2+\alpha^2 - 6 \alpha \beta 
- 36 \beta^2}
{6\alpha \beta H^2[48\beta (\alpha + 3 \beta)H^2+
\alpha + 12\beta] }  \,,\\
\mu_3 &\simeq& -\frac{1728 \beta^2 (\alpha^2 - 15\alpha \beta 
- 24\beta^2) H^6-12\beta (17 \alpha^2 + 120 \alpha \beta 
+ 216 \beta^2) H^4-\alpha (7 \alpha + 36 \beta) H^2
+\alpha - 3 \beta}{12 \alpha \beta H^4 
[48 \beta (\alpha + 3 \beta) H^2+\alpha + 12\beta]} \,,\\
\mu_4 &\simeq& \frac{41472\beta^4 (\alpha + \beta) H^6
-96 \beta^2 (\alpha^2 + 33 \alpha \beta + 36\beta^2)H^4
+6\beta (\alpha^2 - 38 \alpha \beta - 72 \beta^2)H^2
-\alpha (\alpha - 3\beta)}{24 \alpha \beta^2 H^4
[48 \beta (\alpha + 3 \beta) H^2+\alpha + 12\beta]} \,,
\label{mu4}
\ea
and 
\ba
\nu_1 &\simeq& \frac{36 \beta(\alpha + \beta)H^2
+2\alpha+3\beta}{12\beta (\alpha+3\beta)H^2}\,,
\label{nu1}\\
\nu_2 &\simeq& -\frac{2\beta (\alpha-12 \beta)H^2
-\alpha-\beta}{2\alpha \beta H^2} \,,\\
\nu_3 &\simeq& -\frac{144 \beta^2 (\alpha^2 
-15\alpha \beta - 24 \beta^2)H^4 
+ 4\beta(2\alpha^2 -15\alpha \beta - 36\beta^2)H^2 
- 2\alpha^2 - 7\alpha \beta - 6\beta^2}
{48 \alpha \beta^2 (\alpha+3\beta)H^4}\,,\\
\nu_4 &\simeq& \frac{864\beta^2(\alpha + \beta)H^4
+72\beta (\alpha + \beta)H^2+2\alpha 
+ 3\beta}{24\alpha \beta (\alpha+3\beta)H^4}\,.
\label{nu4}
\ea
Let us first consider the case in which the inequality
\be
\beta H^2 \gg 1
\label{bcon}
\ee
is satisfied during inflation.
Then, Eqs.~(\ref{mu1})--(\ref{mu4}) and 
Eqs.~(\ref{nu1})--(\ref{nu4}) approximately reduce to 
\be
\mu_1 \simeq \nu_1 \simeq 
\frac{3(\alpha+\beta)}{\alpha+3\beta}\,,
\qquad
\mu_2 \simeq \nu_2 \simeq
-\frac{\alpha-12 \beta}{\alpha}\,,
\qquad 
\mu_3 \simeq \nu_3 \simeq
-\frac{3 (\alpha^2 - 15\alpha \beta 
- 24\beta^2)}{\alpha (\alpha+3\beta)}\,,
\qquad 
\mu_4 \simeq \nu_4 \simeq
\frac{36 \beta (\alpha+\beta)}
{\alpha (\alpha+3\beta)}\,.
\label{muiap}
\ee
On using these approximate coefficients and 
exploiting the approximation that $H$ is 
constant during inflation, the solutions 
to Eqs.~(\ref{Psieq}) and (\ref{Phieq}) 
in the super-Hubble regime $k/(aH) \ll 1$ 
are given by 
\ba
& &
\Psi=C_1 e^{-Ht}+C_2 e^{-\frac{3(\alpha+\beta)}
{\alpha+3\beta}Ht}
+C_3 e^{(1+\sqrt{1-48 \beta/\alpha})Ht/2}
+C_4 e^{(1-\sqrt{1-48 \beta/\alpha})Ht/2}\,,
\label{Psiso}\\
& &
\Phi=D_1 e^{-Ht}+D_2 e^{-\frac{3(\alpha+\beta)}
{\alpha+3\beta}Ht}
+D_3 e^{(1+\sqrt{1-48 \beta/\alpha})Ht/2}
+D_4 e^{(1-\sqrt{1-48 \beta/\alpha})Ht/2}\,,
\label{Phiso}
\ea
where $C_i$'s and $D_i$'s are constants. 
For $\alpha>0$ and $\beta>0$, the third and fourth 
terms in Eqs.~(\ref{Psiso}) and (\ref{Phiso}) 
grow in time, while the first and second terms decay. 
Depending on the values of $\alpha$ and $\beta$, 
the amplitudes of $\Psi$ and $\Phi$ increase as 
\ba
& &
\{ |\Psi|, |\Phi| \} 
\propto e^{Ht/2}\qquad {\rm for} \quad
\alpha<48 \beta\,,\label{Phiso1} \\
& & 
\{ |\Psi|, |\Phi| \} 
\propto e^{(1+\sqrt{1-48 \beta/\alpha})Ht/2}
\qquad {\rm for} \quad
\alpha>48 \beta\,.
\label{Phiso2} 
\ea
In the coupling range (\ref{Phiso1}), $\Psi$ and $\Phi$ exhibit oscillations with the growing amplitudes. 
In the other coupling regime (\ref{Phiso2}), $\Psi$ and $|\Phi|$ increase even faster than $e^{Ht/2}$. In the large Weyl coupling limit $\alpha \gg 48 \beta$, the gravitational potentials grow rapidly in proportion to $e^{Ht}$.

The above results are valid for $\beta H^2$ exceeding 
the order 1 during inflation. In particular, for the 
couplings in the ranges $\alpha \ll \beta$ and 
$\beta H^2 \gtrsim {\cal O}(1)$, 
the solutions to $\Psi$ and $\Phi$ correspond to 
the limits $\beta/\alpha \to \infty$ in 
Eqs.~(\ref{Psiso}) and (\ref{Phiso}).
This means that even a small Weyl coupling 
constant $\alpha$ induces the exponential growth 
of gravitational potentials. 
Then, the homogeneous FLRW background is violated 
by the rapid growth of $\Psi$ and $\Phi$ in the 
perturbed metric (\ref{Newtonian}). 
In the above analytic estimation we used the approximation $\beta H^2 \gg 1$, but in Sec.~\ref{numesec}, we will show that, even for $\beta H^2={\cal O}(1)$, both the amplitudes of 
$\Psi$ and $\Phi$ increase exponentially.
The fact that $\Psi$ and $\Phi$ are subject to exponential growth does not depend on the gauge choices either. 
Since all scalar perturbations are always determined through two independent modes, at least one of them needs to be unstable. 
It should be also pointed out that this instability is purely classical.

Besides $\Psi$ and $\Phi$, there are also other 
gauge-invariant scalar perturbations. 
Let us consider the evolution of the curvature 
perturbation $\zeta$ defined in Eq.~(\ref{zeta}). 
In the Newtonian gauge, we will describe a method to find 
a closed differential equation for $\zeta$, where 
$\zeta$ is given by Eq.~\eqref{zeta2}.
We can think of Eq.~(\ref{zeta2}) as an equation that sets, on shell, $\zeta$ as a function of the other fields. Therefore, we proceed by adding a term to the second-order scalar Lagrangian density $L_s$, as follows:
\begin{equation}
\bar{L}_{s} = L_{s}+b_1(t)\left\{
\Phi-\frac{H}{\dot{R}} \left[ 
6 (\ddot{\Phi}+4H \dot{\Phi})+\frac{4k^2}{a^2} 
\Phi-6H \dot{\Psi}
-12(2H^2+\dot{H})\Psi+\frac{2k^2}{a^2} \Psi
\right]-\zeta\right\}^2\,,
\end{equation}
where $b_1(t)$ is a function of $t$.
It is clear that at this level $\zeta$ is just a Lagrange multiplier, 
and its equation of motion, algebraic for $\zeta$ itself, makes 
$\bar{L}_{s}$ reduce to the original Lagrangian density $L_{s}$. 
We choose the coefficient $b_1(t)$ to cancel the term in 
$\dot{\Psi}^2$, and, by doing so, also the term in 
$\ddot{\Phi}^2$ cancels out from the Lagrangian.

After a few integrations by parts, we see that the field $\Psi$ 
can be set to be a Lagrangian multiplier, and as such, it is 
integrated out from the Lagrangian by using its equation of motion. 
By doing so, we arrive at an equivalent Lagrangian density, 
$\bar{L}_{s}=\bar{L}_{s}(\dot{\zeta},\dot{\Phi},\zeta,\Phi)$, 
which then depends on two propagating d.o.f.s, 
$\Phi$ and $\zeta$, as expected. 
The reduced Lagrangian density contains the products 
of kinetic terms of the form 
\be
\bar{L}_s \supset 
\bar{K}_{11}\dot{\zeta}^2
+\bar{K}_{22} \dot{\Phi}^2
+2\bar{K}_{12}\dot{\zeta} \dot{\Phi}\,.
\ee
For positive values of $\alpha$ and $\beta$, 
we have $\bar{K}_{11}>0$ and 
$\bar{K}_{11}\bar{K}_{22}-\bar{K}_{12}^2<0$
in the slow-roll limit. 
Hence there is one ghost mode besides the other no-ghost mode. This property 
agrees with the no-ghost conditions 
derived in Sec.~\ref{ngsec} for the flat gauge. 

To derive the closed-form perturbation equation of $\zeta$, we proceed as follows. From the equation of motion for 
the field $\Phi$, which we write in the form ${\cal E}_{\Phi}=0$, 
we find an expression for $\ddot{\Phi}$ that can be inserted into 
the equation for $\zeta$, i.e., ${\cal E}_{\zeta}=0$.
Now, we take the time derivative of this last equation to 
obtain $\dot{\cal E}_\zeta=0$. 
We can still substitute this new equation into the expression of $\ddot{\Phi}$, previously found, and solve it for $\dot{\Phi}$. 
We repeat the step on considering now the equation $\ddot{\cal E}_\zeta=0$, and after replacing it with the two expressions for $\ddot{\Phi}$ and $\dot{\Phi}$, we can solve it with respect to $\Phi$ itself. 
At this point, we replace all these $\Phi$-related expressions into the equation of motion ${\cal E}_\zeta=0$. Then, we obtain the 
fourth-order differential equation 
\be
\ddddot{\zeta}+\lambda_1 H \dddot{\zeta}+\lambda_2 H^2 
\ddot{\zeta}+\lambda_3 H^3 \dot{\zeta}+\lambda_4 H^4 \zeta=0\,,
\label{zetaeq1}
\ee
where $\lambda_i$'s are time-dependent dimensionless 
coefficients. 

Taking the sub-Hubble limit $k/(aH) \gg 1$ with 
$\epsilon \to 0$, the coefficients in Eq.~(\ref{zetaeq1}) 
reduce to 
\be
\lambda_1 \simeq \frac{168 \beta H^2+1}{12\beta H^2}\,,
\qquad
\lambda_2 \simeq \frac{2k^2}{(aH)^2}\,,
\qquad 
\lambda_3 \simeq \frac{(120 \beta H^2+1)k^2}
{12 \beta H^4 a^2}\,, \qquad
\lambda_4 \simeq 
\frac{k^4}{(aH)^4}\,.
\ee
Using the solution $\zeta=\zeta_0 e^{-i \int \omega \rd t}$ under the WKB approximation $\omega=c_s k/a \gg H$, the leading-order dispersion relation is given by $\omega^4-2k^2 \omega^2/a^2+k^4/a^4 \simeq 0$.
Hence the curvature perturbation propagates with the luminal speed for the modes deep inside the Hubble radius.

For super-Hubble perturbations, we take the limits $k/(aH) \to 0$ 
and $\epsilon \to 0$ for the coefficients $\lambda_i$. 
For the coupling in the range $\beta H^2 \gg 1$, 
we approximately have
\be
\lambda_1 \simeq 
\frac{8(8\alpha+3 \beta)}{5\alpha+3\beta}\,,\qquad
\lambda_2 \simeq \frac{157 \alpha^2 
+111\alpha \beta + 36 \beta^2}
{\alpha(5\alpha+3\beta)}\,,\qquad
\lambda_3 \simeq \frac{2(\alpha+6\beta)
(49\alpha+15\beta)}
{\alpha(5\alpha+3\beta)}\,,\qquad 
\lambda_4 \simeq 0\,.
\ee
In this regime, we can integrate Eq.~(\ref{zetaeq1}) to give
\be
\zeta= 
c_{1} +c_{2} e^{-\frac{( 49 \alpha+15 \beta ) Ht}
{5 \alpha+3 \beta}}+c_{3} 
e^{(-3 +\sqrt{1-48 \beta/\alpha}) H t/2}
+c_{4}e^{-(3 +\sqrt{1-48 \beta/\alpha}) H t/2}\,.
\label{zesol}
\ee
For positive values of $\alpha$ and $\beta$, the last three 
terms in Eq.~(\ref{zesol}) decay in time. 
Hence, the curvature perturbation approaches a constant 
$c_1$ after the Hubble radius crossing.
In the above estimation, we have used the approximation $\beta H^2 \gg 1$, but we have numerically confirmed that $\zeta$ approaches a constant even for $\beta H^2={\cal O}(1)$.

Despite the exponential increase of two gravitational potentials during inflation, there is a particular gauge-invariant combination $\zeta$ that does not grow in the large-scale limit. 
Unlike the standard single-field slow-roll inflation, however, we have two propagating d.o.f.s in the scalar sector. 
In the description of Lagrange multipliers explained above, the two dynamical d.o.f.s correspond to the perturbations $\Phi$ and $\zeta$. Even though $\zeta$ is not enhanced after the Hubble radius crossing, the other dynamical field $\Phi$ is subject to exponential growth. 
Thus, the analysis in the Newtonian gauge shows that the Weyl curvature term violates the homogeneous inflationary background.

\subsection{Flat gauge}
\label{flatsec}

In the flat gauge with the gauge conditions $\psi=0$ and $E=0$, we have ${\cal A}=A$ and ${\cal B}=aB$ in Eq.~(\ref{eq:FG}). Then, the perturbed line element is given by 
\be
\rd s^2=-\left(1+2{\cal A} \right) \rd t^2
+2 \partial_i {\cal B}  \rd t \rd x^i
+a^2 (t) \delta_{ij} \rd x^i \rd x^j\,.
\label{metflat}
\ee
On the expanding cosmological background ($H \neq 0$), 
the coordinate transformation vector $\xi^{\mu}$ 
is always regular for the flat gauge.
The gauge-invariant variables ${\cal A}$ and ${\cal B}$ 
are related to $\Psi$ and $\Phi$ according to 
\be
{\cal A}=\Psi-\frac{\rd}{\rd t} 
\!\left( \frac{\Phi}{H} \right) \,,
\qquad 
{\cal B}=\frac{\Phi}{H}\,.
\label{AB}
\ee
While ${\cal B}$ is directly proportional to $\Phi$, 
${\cal A}$ corresponds to a combination of $\Psi$ and $\Phi$.

After setting $\psi=0=E$ in the perturbation equations 
of motion, the dynamical system in the flat gauge has 
two propagating d.o.f.s ${\cal A}$ and ${\cal B}$ (or $\Phi$).
To derive the closed differential equation for ${\cal A}$, 
we solve the two equations ${\cal E}_{A}=0$ and ${\cal E}_{B}=0$ 
for $\ddot{\cal B}$ and $\dot{\cal B}$. 
Following a similar procedure to that performed in the Newtonian gauge, we can express the terms $\dot{\cal B}$ and ${\cal B}$ in terms of ${\cal A}$ and its derivatives.
Taking the time derivative of the ${\cal B}$ equation and combining it with the $\dot{\cal B}$ equation, we obtain the fourth-order differential equation 
of ${\cal A}$ in the form 
\be
\ddddot{\cal A}+\tau_1 H \dddot{\cal A}
+\tau_2 H^2 \ddot{\cal A}+\tau_3 H^3 \dot{\cal A}
+\tau_4 H^4 {\cal A}=0\,,
\label{Aeq}
\ee
where $\tau_i$'s are time-dependent functions.

Taking the sub-Hubble limit $k \gg aH$ with $\epsilon \to 0$, 
the coefficients in Eq.~(\ref{Aeq}) reduce to 
\be
\tau_1 \simeq 10\,,
\qquad
\tau_2 \simeq \frac{2k^2}{(aH)^2}\,,
\qquad 
\tau_3 \simeq \frac{6k^2}{(aH)^2}\,, \qquad
\tau_4 \simeq 
\frac{k^4}{(aH)^4}\,.
\ee
On using the WKB approximation, it follows that the 
field ${\cal A}$ propagates with the speed of light.

For super-Hubble modes $k \ll aH$ with the coupling 
$\beta H^2 \gg 1$, taking the slow-roll limit 
$\epsilon \to 0$ gives
\be
\tau_1 \simeq 6\,,
\qquad
\tau_2 \simeq \frac{11 \alpha+12\beta}{\alpha}\,,
\qquad 
\tau_3 \simeq \frac{6(\alpha+6 \beta)}{\alpha}\,, \qquad
\tau_4 \simeq -\frac{\alpha+6 \beta}{3 \alpha  \beta H^2}\,.
\ee
In this regime, Eq.~(\ref{Aeq}) can be integrated to give
\begin{align}
{\cal A} &=c_1 e^{-(9\alpha\beta H+\sqrt{\Delta_1-12 \alpha\beta H\sqrt{\Delta_2}})t/(6\alpha\beta)}
+c_2 e^{-(9\alpha\beta H-\sqrt{\Delta_1-12 \alpha\beta H\sqrt{\Delta_2}})t/(6\alpha\beta)} 
\nonumber \\
&~~~+c_3 e^{-(9\alpha\beta H+\sqrt{\Delta_1+12 \alpha\beta H\sqrt{\Delta_2}})t/(6\alpha\beta)}
+c_4e^{-(9\alpha\beta H-\sqrt{\Delta_1+12 \alpha\beta H\sqrt{\Delta_2}})t/(6\alpha\beta)}\,,
\label{Aso1}
\end{align}
where
\ba
\Delta_1 &\equiv&
9 H^2 \alpha \beta^2 (5 \alpha - 24\beta)\,,\\
\Delta_2 &\equiv& 
3 \beta (\alpha+6\beta) \left[ 
3\beta H^2 (\alpha+6\beta)
+\alpha \right]\,.
\ea
The amplitudes of the first two terms in Eq.~(\ref{Aso1}) 
decrease in proportion to $e^{-3Ht/2}$, while 
the third term decreases as $\propto e^{-3Ht}$. 
Taking the limit $\beta H^2 \gg 1$, the leading-order 
contribution to the term 
$\Delta_1+12 \alpha\beta H\sqrt{\Delta_2}$ is 
$(9 \alpha \beta H)^2$. Then, for $\alpha>0$ and 
$\beta>0$, the leading-order contribution to the last term 
in Eq.~(\ref{Aso1}) is the constant $c_4$.
Picking up the next-to-leading 
correction, we obtain the following solution
\be
\mathcal{A} \simeq c_4 \left( 
1+\frac{t}{18 \beta H} \right)\,.
\label{Afsol}
\ee
For the number of $e$-foldings $N$ of order 10, the correction 
induced by the time-dependent terms in Eq.~(\ref{Afsol}) is 
suppressed compared to the leading-order constant term.

Following a similar procedure performed for the perturbation ${\cal A}$, we can also derive 
the fourth-order differential equation for ${\cal B}$ in the form 
\be
\ddddot{\cal B}+\eta_1 H \dddot{\cal B}
+\eta_2 H^2 \ddot{\cal B}+\eta_3 H^3 \dot{\cal B}
+\eta_4 H^4 {\cal B}=0\,,
\label{Beq}
\ee
where the $\eta_i$'s are time-dependent coefficients and are not identical to the $\nu_i$'s in Eq.~(\ref{Phieq}). 
If we use the variable $\Phi$ instead of ${\cal B}$, 
the coefficients of the fourth-order differential equation for $\Phi$ exactly coincide 
with those derived in the Newtonian gauge.

Using the WKB approximation for the modes deep inside the Hubble radius, the perturbation ${\cal B}$ obeys 
\be
\ddddot{\cal B}+\frac{2k^2}{a^2} \ddot{\cal B}
+\frac{k^4}{a^4}{\cal B} \simeq 0\,,
\label{Bsubeq}
\ee
so that ${\cal B}$ propagates with the speed of light.
Taking the super-Hubble limit ($k \ll aH$) with $\epsilon \to 0$, the coefficients in Eq.~(\ref{Beq}) reduce to 
\ba
\eta_1 &\simeq& \frac{36(\alpha+\beta)\beta H^2
+2\alpha+3\beta}{12(\alpha+3\beta)\beta H^2}\,,\label{eta1} 
\\
\eta_2 &\simeq& \frac{2(12\beta-\alpha)H^2+1}
{2\alpha H^2}\,,\\
\eta_3 &\simeq& -\frac{72 \beta 
(\alpha^2 - 15 \alpha \beta - 24\beta^2)H^4
-2(\alpha^2 + 42\alpha \beta + 36 \beta^2)H^2
-2\alpha-3\beta}{24 \alpha \beta (\alpha+3\beta)H^4}\,,\\
\eta_4 &\simeq& \frac{5184 \beta^3 (\alpha+\beta)H^4
+12\alpha \beta (\alpha+9\beta)H^2
+\alpha (2\alpha+3\beta)}
{144 \alpha \beta^2 (\alpha+3\beta)H^4}\,.
\label{eta4}
\ea
For $\beta H^2 \gg 1$, the coefficients $\eta_{1,2,3,4}$ approximately reduce to the values $\nu_{1,2,3,4}$ given in Eq.~(\ref{muiap}), respectively. 
The field ${\cal B}$ is subject to exponential growth during inflation analogous to $\Phi$, see Eqs.~(\ref{Phiso1})--(\ref{Phiso2}). 
This means that, even though ${\cal A}$ does not grow significantly, the other metric perturbation ${\cal B}$ in the line element (\ref{metflat}) increases rapidly to violate the FLRW background. 

In the flat gauge, the perturbation $\zeta$ is given by 
\be
\zeta =-\frac{H}{\dot R}\,\delta R=
-\frac{H}{\dot R} \left[\frac{2 k^{2}
\dot{\mathcal{B}}} {a^{2}}
-6 H \dot{\mathcal{A}}
+\left(\frac{2 k^{2}}{a^{2}}-24 H^{2}-12 \dot{H}\right) 
\mathcal{A}+\frac{4 \mathcal{B} H \,k^{2}}{a^{2}}\right]\,.
\label{eq:zeta_FG}
\ee
To derive the fourth-order differential equation for $\zeta$, we build the following Lagrangian density
\be
\bar{L}_s = L_s
+ b_2(t)\left\{\zeta+\frac{H}{\dot R}\left[  
\frac{2 k^{2} \dot{\mathcal{B}}} {a^{2}}
-6 H \dot{\mathcal{A}}
+\left(\frac{2 k^{2}}{a^{2}}-24 H^{2}-12 \dot{H}\right) 
\mathcal{A}+\frac{4 \mathcal{B} H \,k^{2}}{a^{2}} \right]\right\}^2\,,
\ee
which is equivalent to the original Lagrangian density $L_s$. 
The function $b_2(t)$ needs to be chosen so that the kinetic 
term of $\mathcal{A}$ vanishes identically. 
After a few integrations by parts, the field $\mathcal{A}$ becomes a Lagrange multiplier which can be integrated out, 
leaving $\zeta$ and $\mathcal{B}$ as two dynamical d.o.f.s. 
After varying $\bar{L}_s$ with respect to $\zeta$ and $\mathcal{B}$, 
we can proceed along the same lines as finding the equations of motion 
for the fields $\mathcal{A}$ and ${\cal B}$.
This leads to the closed differential equation for $\zeta$ 
in the form (\ref{zetaeq1}), where $\lambda_i$'s are exactly 
the same as those derived in the Newtonian gauge. 
Hence $\zeta$ approaches a constant after the Hubble radius 
crossing. However, the fact that the other dynamical 
perturbation ${\cal B}$ grows exponentially means that 
the instability of the FLRW background cannot be avoided.
We also note that the fourth-order differential equation 
of $\Psi$ exactly coincides with the one obtained in the Newtonian 
gauge. Hence the two gravitational potentials $\Psi$ and $\Phi$ 
are unstable in the flat gauge as well, by reflecting the fact 
that both $\Psi$ and $\Phi$ contain the dependence of $aB$.

\subsection{Unitary gauge}

Let us finally discuss the evolution of scalar perturbations 
in the unitary gauge with $\delta R=0$. 
Since the curvature perturbation $\zeta$ is equivalent 
to $\psi$, the gauge condition translates to 
\be
\delta R=6 \left( \ddot{\zeta}+4H \dot{\zeta} 
\right)+\frac{4k^2}{a^2}\zeta-6H \dot{A}
-12\left( 2H^2+\dot{H} \right)A+\frac{2k^2}{a^2}A+
\frac{2k^2}{a} \left( \dot{B}+3HB \right)=0\,.
\label{delR1}
\ee

Using this condition together with the perturbation 
equations of motion, we can derive the fourth-order 
differential equation for $\zeta$. 
We first solve the perturbation equation 
${\cal E}_A=0$ for $\ddot{B}$ and take the time 
derivative of Eq.~(\ref{delR1}) to obtain the 
first derivative $\dot{B}$. Using 
Eq.~(\ref{delR1}) to solve for $\dot{B}$, 
one can express $B$ and its time derivatives 
in terms of $\zeta, A$, and their time derivatives. 
In this way, all the $B$-dependent quantities can 
be eliminated from the perturbation equations 
of motion. The next step is to remove the $A$-dependent 
terms. On using the two equations ${\cal E}_{\psi}=0$ 
and ${\cal E}_{E}=0$, we can solve for $\dddot{A}$ and 
$\ddot{A}$. Then, following a similar procedure as before, 
it is possible to express $A$ in terms of the derivatives of 
$\zeta$ up to third order.
Taking the time derivative of this equation and eliminating 
the $\dot{A}$ term, we obtain the fourth-order differential 
equations of $\zeta$ with the exactly same coefficients as $\lambda_{1,2,3,4}$ in Eq.~(\ref{zetaeq1}).
Then, the constancy of $\zeta$ after the Hubble radius 
crossing also holds in the unitary gauge. 
This result is consistent with the analysis 
of Ref.~\cite{Deruelle:2010kf} in the Einstein frame. 
Similarly, we obtain the same fourth-order differential equation 
for ${\cal A}$ as Eq.~(\ref{Aeq}), so the solution in the super-Hubble regime is 
given by Eq.~(\ref{Afsol}).

The closed differential equations for $\Psi$ and $\Phi$ 
can be also obtained by using the following relations
\be
A=\Psi-\dot{\cal B}_u\,,\qquad 
\zeta=\Phi-H{\cal B}_u\,,
\ee
where 
\be
{\cal B}_u \equiv aB\,.
\ee
The gauge condition (\ref{delR1}) and the perturbation equations 
of motion can be now expressed in terms of the gauge-invariant 
variables $\Psi$, $\Phi$, ${\cal B}_u$ and their time derivatives. 
Indeed, the above change of variables automatically removes 
the ${\cal B}_u$-dependent terms from the two perturbation 
equations ${\cal E}_A=0$ and ${\cal E}_B=0$. 
Combining these two, it is straightforward to 
derive the fourth-order differential equations for
$\Psi$ and $\Phi$. Again, we find that they are identical to Eqs.~(\ref{Psieq}) and (\ref{Phieq}) derived in the Newtonian gauge, respectively, with the completely same coefficients.
Hence the same instabilities of $\Psi$ and $\Phi$ 
are present after the Hubble radius crossing, while 
the growth of $\zeta$ and ${\cal A}$ is suppressed.

In terms of the gauge-invariant variables, the perturbed line element in the unitary gauge can be expressed as
\be
\rd s^2=-\left[ 1+2{\cal A}+2 \frac{\rd}{\rd t} 
\left( \frac{\zeta}{H} \right) \right] \rd t^2
+\frac{2}{H} \left( \partial_i \Phi-\partial_i \zeta 
\right)\rd t \rd x^i+a^2(t)(1+2\zeta) 
\delta_{ij}\rd x^i \rd x^j\,.
\label{unimet}
\ee
Due to the suppressed growth of ${\cal A}$ and $\zeta$, 
metric perturbations in the $g_{00}$ and $g_{ij}$ 
components are not subject to classical instabilities. 
However, the $\partial_i \Phi$ term in the $g_{0i}$ 
component exhibits an exponential increase after 
the Hubble radius crossing. Since the $\partial_i \zeta$ term 
approaches a constant in the super-Hubble regime, 
the dominance of $\partial_i \Phi$ 
over $\partial_i \zeta$ in $g_{0i}$ 
leads to the instability of the FLRW background. 
Indeed, we numerically confirmed that 
the gauge-invariant perturbation ${\cal B}_u=(\Phi-\zeta)/H$ 
grows exponentially in the super-Hubble regime due to 
the enhancement of $\Phi$.
We have thus analytically shown that, for any physical gauge choices, the Weyl curvature makes the inflationary FLRW background unstable.

\section{Numerical simulations with the 
discussion of initial conditions}
\label{sec:numerics}

In this section, we will numerically confirm the instability of 
the FLRW background in Weyl gravity with the $\beta R^2$ term. 
For this purpose, we first discuss the choice of initial 
conditions of perturbations and then proceed to the 
numerical analysis.

\subsection{Fourth-order system and 
initial conditions}
\label{fouthsec}

We have learned so far that each of the considered perturbation fields obeys a fourth-order differential equation, which, under the WKB approximation, is solved as the solutions describing waves 
propagating with the speed of light. 
We will show that it is indeed possible, starting from 
the reduced action of a single scalar field $v$ possessing the term ${\ddot v}^2$, to find an equivalent Lagrangian density of 
two scalar fields with second-order equations of motion. 
The discussion in this section can be applied to any dynamical perturbation $v$ with some gauge choices, 
but in Sec.~\ref{Inisec} we will consider the flat gauge for concreteness.

For the modes deep inside the Hubble radius, we should expect 
the field $v$ in Fourier space to satisfy a fourth-order equation of motion, which can be derived by the following approximate Lagrangian
\begin{equation}
L_v \simeq Q \left( {\ddot v}^2 -2\frac{k^2}{a^2}\,{\dot v}^2+\frac{k^4}{a^4}\,v^2 \right)\,.
\label{eq:lag4th}
\end{equation}
Here and in the following, we assume that the function $Q=Q(t,k^2)$ 
can be either positive or negative. 
In Sec.~\ref{Inisec}, we will see that 
it is possible 
to obtain a Lagrangian density\footnote{The Lagrangian, in this case, 
will take the following general form $L_v=Q\,(\ddot{v}^2-2Q_1\dot{v}^2+Q_2 v^2)$, 
where $Q$, $Q_1$, and $Q_2$ are functions of time and $k^2$.} 
reducing to the form \eqref{eq:lag4th} 
in the high-$k$ regime with $v$ 
related to the perturbations 
${\cal A}$ and ${\cal B}$. 
We note that, in the action (\ref{eq:lag4th}), we are assuming 
the high momentum regime in which the WKB approximation holds 
for the dynamics of $v$.

Then, we can introduce an auxiliary field, $w$, as
\begin{equation}
L_v\simeq Q \left[{\ddot v}^2 -2\frac{k^2}{a^2}\,{\dot v}^2+\frac{k^4}{a^4}\,v^2 -(b_2 w+\ddot v)^2 \right]\,,
\label{eq:lag4th_2}
\end{equation}
where $b_2(t,k^2)$ is a general function, so far undetermined. 
By integrating out the field $w$, we find once more the original Lagrangian density. Therefore, the two Lagrangian densities \eqref{eq:lag4th} and~\eqref{eq:lag4th_2} are equivalent to each other, both of which lead to the same dynamics. 
It is also clear that the term in $\ddot{v}^2$ vanishes 
in Eq.~\eqref{eq:lag4th_2} for the new Lagrangian density. 
At this level, we can introduce the 
quantity
\begin{equation}
v=v_2+\frac{a^2b_2}{2k^2}\,w\,,
\end{equation}
which is meant to diagonalize 
the kinetic matrix. 
Then, we perform the other field redefinitions
\begin{equation}
v_2=\frac{a}{k}\,v_3\,,\qquad 
w = \frac{k}{a}\,w_3\,,
\end{equation}
as to make the kinetic terms only background dependent, i.e.,\ 
independent of the wave number $k$.

Having assumed that $Q\neq0$, we can further introduce the following field redefinitions
\begin{equation}
v_3=\frac{a^{3/2}}{2\sqrt{|Q|}}\,v_4\,,\qquad
w_3=\frac{a^{3/2}}{b_2\sqrt{|Q|}}\,w_4\,,
\end{equation}
to obtain canonical kinetic terms 
for the fields $v_4$ and $w_4$.
At this level, the Lagrangian for the modes deep 
inside the Hubble radius reduces to
\begin{equation}
L_v\simeq -\mathrm{sign}(Q)\left[\frac{a^3}{2}\,(\dot{v}_4^2-\dot{w}_4^2)-\frac{k^2a}{4}\,(v_4^2-3w_4^2+2v_4w_4)\right].
\end{equation}
This Lagrangian still leads to the dynamics of perturbations 
propagating with the speed of light. 
However, the effective mass matrix $C=\tfrac14\,k^2a\,\begin{psmallmatrix}
1 & 1 \\
1 & -3   
\end{psmallmatrix}$ cannot be diagonalized by any real (finite and constant-in-time) Lorentz transformation that would leave instead the kinetic matrix in the canonical form. Therefore, 
the two modes $v_4$ and $w_4$ are not completely decoupled.

Nonetheless, we can use the WKB approximation and look for solutions of the kind $v_4\propto e^{-i\int \omega {\rm d}t}$ 
and $w_4\propto e^{-i\int \omega {\rm d}t}$, for which $\ddot{v}_4\propto -\omega^2 v_4$ and $\ddot{w}_4\propto -\omega^2 w_4$. In this case, it follows that the dispersion relation $\omega=k/a$ needs to hold, which means that both modes 
propagate with the luminal speeds. 
Furthermore, we obtain the relation $w_4=v_4$, and hence the dynamics of the mode $w_4-v_4$ is set to vanish. 
Still, proper initial conditions need to be imposed 
on the field $v_4$ (or $w_4$).

To better understand the behavior of solutions or the choice of initial conditions, we use the perturbation equations 
of motion in the high-$k$ regime expressed in terms of 
the conformal time $\eta=\int a^{-1}{\rm d}t$,
\begin{equation}
\frac{\rd^2 v_4}{\rd \eta^2}
\simeq-\frac{k^2}2\,(v_4+w_4)\,,\qquad
\frac{\rd^2 w_4}{\rd \eta^2} 
\simeq \frac{k^2}2\,(v_4-3w_4)\,.
\end{equation}
Then, we obtain the following general solutions
\begin{align}
w_4 &= c_1\sin(k\eta)+c_2\cos(k\eta) 
+ c_3\,k\eta\sin(k\eta) + c_4\, k\eta \cos(k\eta)\,,\\
v_4 &= c_1\sin(k\eta)+c_2\cos(k\eta) 
+c_3\, [k\eta \sin(k\eta) +4\cos(k\eta)] + c_4\, [k\eta \cos(k\eta) -4\sin(k\eta)]\,,
\end{align}
where $c_1$, $c_2$, $c_3$, and $c_4$ are integration constants. 
The choice of exact plane wave initial conditions corresponds to $c_3=0=c_4$, so that $v_4=w_4$. 
As in the usual normalization scheme of the Bunch-Davies vacuum,\footnote{From a purely classical point of view, the initial condition representing a plane wave consists of setting 
$\rd v_4/\rd \eta=-ik v_4$, and so on for all other higher derivatives. The normalization of $v_4$ itself is not established as the field $C v_4$ is still a solution of the equations of motion ($C$ is a constant). 
In this case, the physical quantity to consider 
is $v_4(\eta)/v_4(\eta_i)$.}
we choose a positive frequency solution 
$v_4=w_4={\cal C}_0 e^{-i k \eta}$
and impose the conditions $\sqrt{|v_4^2||(a^3\dot{v}_4)^2|}=1/2$ and 
$\sqrt{|w_4^2||(a^3\dot{w}_4)^2|}=1/2$.
This fixes the coefficient ${\cal C}_0$ to be $1/(a\sqrt{2k})$, 
and hence 
\begin{equation}
v_4=w_4=
\frac{1}{a\sqrt{2k}}\, e^{-i \int \frac{k}{a}{\rm d}t}\,.
\end{equation}
In terms of the original perturbation $v$, we have 
\be
v=\frac{a^{5/2}}{2k \sqrt{|Q|}} \left( v_4+w_4 \right)
=\frac{a^{3/2}}{\sqrt{2k^3 |Q|}} 
e^{-i \int \frac{k}{a}{\rm d}t}\,.
\ee
We will choose this as the initial condition of $v$ 
for the modes deep inside the Hubble radius. 
Note that we will not discuss the quantization of perturbations in our theory. 
We have already shown analytically the presence of violent classical instabilities for the modes after the Hubble radius crossing. In other words, the scalar ghost is not of the soft type in our theory. 
The classical instabilities induced by the ghost make the quantization of perturbations irrelevant. Only for a stable classical background, it would be worth investigating the quantization procedure.

\subsection{Initial conditions for the perturbations 
${\cal A}$ and ${\cal B}$}
\label{Inisec}

In Sec.~\ref{fouthsec}, we assumed the existence of 
the Lagrangian density \eqref{eq:lag4th} leading 
to the closed fourth-order differential equation. 
In this section, we will prove its existence 
by considering perturbations in the flat gauge.
For this purpose, we will proceed as follows. 
In the flat gauge, the kinetic Lagrangian for the fields $\mathcal{A}$ and $\mathcal{B}$ was already 
discussed in Eq.~\eqref{LsK}. 
We remind the reader about the field redefinitions that were introduced 
to obtain canonical kinetic terms. 
The field redefinitions given in Eqs.~(\ref{fre1}) and (\ref{B2def}) allow us to obtain the Lagrangian density 
of two canonically normalized fields 
${\cal A}_2$ and ${\cal B}_2$ in the form 
(\ref{cano}). The original perturbations 
${\cal A}$ and ${\cal B}$ are related to 
${\cal A}_2$ and ${\cal B}_2$, as
\be
{\cal A}=\frac{1}{6 \Mpl H} \left( 
\frac{{\cal A}_2}{\sqrt{\beta}}
+\frac{\sqrt{3} {\cal B}_2}{\sqrt{\alpha}} 
\right)\,,\qquad 
{\cal B}=\frac{\sqrt{3} a^2 {\cal B}_2}
{2\Mpl k^2 \sqrt{\alpha}}\,.
\label{ABd}
\ee

{}From the canonical expression (\ref{cano}), we wish to find the Lagrangian density of the form (\ref{eq:lag4th}).
We perform the following field redefinitions
\begin{equation}
\mathcal{A}_2 = \frac{1}{\sqrt2}\,(\mathcal{Y}-\mathcal{Z})\,,\qquad
\mathcal{B}_2 = \frac{1}{\sqrt2}\,(\mathcal{Y}+\mathcal{Z})\,,
\label{AB2d}
\end{equation}
so that the kinetic terms are simplified to
\begin{equation}
L_{s} \ni -a^3 \dot{\mathcal{Y}}\dot{\mathcal{Z}}+\dots 
= \mathcal{Z}\,\frac{\rd}{\rd t}(a^3 \dot{\mathcal{Y}}) 
+ \dots\,,
\label{Lso}
\end{equation}
up to total derivatives. 
Varying the full Lagrangian density $L_s$ with respect to $\mathcal{Z}$, 
it follows that $\mathcal{Z}$ can be expressed in terms of 
$\mathcal{Y}$ and its first and second-time derivatives. 
Substituting $\mathcal{Z}$ into the second expression of Eq.~(\ref{Lso}), we find that the Lagrangian density can be expressed in the form 
\begin{equation}
L_s=Q\,(\ddot{\mathcal{Y}}^2-2Q_1\,\dot{\mathcal{Y}}^2
+Q_2 \,\mathcal{Y}^2)\,,
\end{equation}
where $Q$, $Q_1$, and $Q_2$ depend on $t$ and $k^2$.
It should be noticed that, up to this point, we have not made any approximation for particular wavenumbers. 
However, if we look for the behavior of the three quantities $Q$, $Q_1$, and $Q_2$ in the high-$k$ regime, we find
\begin{equation}
Q \simeq \frac{27 \alpha \beta a^{7} H^{2}}{\left(\alpha-3 \beta \right) \left(\sqrt{\alpha}-\sqrt{3\beta}\right)^{2} k^{4}}\,,\qquad
Q_1 \simeq \frac{k^2}{a^2}\,,\qquad
Q_2 \simeq \frac{k^4}{a^4}\,,
\label{Qco}
\end{equation}
which means that\footnote{For the special case where $\alpha=3\beta$, by looking at Eq.\ \eqref{LsK}, we can see that the field $\mathcal{B}$ can be easily integrated out in terms of $\mathcal{A}$ and its first and second time derivatives. Alternatively, we can still use the procedure described here, but now $Q\simeq-a^3/(6H^2)$ and $\mathcal{Z}\simeq ik\mathcal{Y}/(aH)\gg\mathcal{Y}$. } $\mathrm{sign}(Q)=\mathrm{sign}(\alpha-3\beta)$. 
The behavior of $Q_1$ and $Q_2$ makes sure that the propagation of 
the mode ${\cal Y}$ is luminal in the WKB approximation scheme. 

Using the discussion given in Sec.~\ref{fouthsec}, 
${\cal Y}$ plays the role of the field $v$ 
with $Q$ given in Eq.~(\ref{Qco}).
Then, we can choose the initial condition 
of ${\cal Y}$, as 
\be
{\cal Y}=\frac{a^{3/2}}{\sqrt{2k^3 |Q|}} 
e^{-i \int \frac{k}{a}{\rm d}t}
\simeq \frac{k^{1/2} \sqrt{|\alpha-3\beta|}\, 
|\sqrt{\alpha}-\sqrt{3\beta}|}
{3 \sqrt{6\alpha \beta}\,a^2 H} 
\,e^{ik (e^{-N}-1)/(a_iH_i)}\,,
\label{calY}
\ee
where $N=\ln(a/a_i)$ is the $e$-folding number, 
and we set $N=0$ at the initial time.
Substituting this solution into the relation between 
${\cal Z}$ and ${\cal Y}$, $\dot{{\cal Y}}$, 
$\ddot{{\cal Y}}$, we find 
\be
{\cal Z}= \frac{\sqrt{\alpha}+\sqrt{3\beta}}
{\sqrt{\alpha}-\sqrt{3\beta}}\,{\cal Y}
-\frac{6i\sqrt{3\alpha \beta}}
{(\sqrt{\alpha}-\sqrt{3\beta})^2} 
\frac{aH}{k}{\cal Y}
+{\cal O} \left( \frac{a^2H^2}{k^2}\right) {\cal Y} \,,
\label{ZYre}
\ee
which is valid for the modes deep inside the Hubble radius.\footnote{Alternatively, we can write the Lagrangian density in the form $L_s=\mathcal{Y}\,(\rd/\rd t)
(a^3 \dot{\mathcal{Z}})+\cdots$ and 
vary the action with respect to ${\cal Y}$ 
and use the equation of motion for ${\cal Y}$ to express 
$L_s$ with respect to ${\cal Z}$ and its derivatives.
In the high-$k$ regime, the Lagrangian density 
is given by $L_s \simeq 
\tilde{Q}\,[\ddot{\mathcal{Z}}^2-(2k^2/a^2)\,
\dot{\mathcal{Z}}^2+(k^4/a^4)\,\mathcal{Z}^2]$, 
where $\tilde{Q}=27 \alpha\beta \,a^{7} H^{2}
/[(\alpha-3 \beta )(\sqrt{\alpha}+\sqrt{3\beta})^2 k^4]$. 
The WKB solution to ${\cal Z}$ derived by this procedure 
is consistent with the leading-order relation of Eq.~(\ref{ZYre}).}

On using Eqs.~(\ref{ABd}), (\ref{AB2d}), and (\ref{ZYre}), the WKB solutions to ${\cal A}$ and ${\cal B}$ are 
\ba
{\cal A} &=& \frac{1}{6\sqrt{2}\Mpl H} 
\left[ \frac{1}{\sqrt{\beta}} ({\cal Y}-{\cal Z})
+\sqrt{\frac{3}{\alpha}} ({\cal Y}+{\cal Z}) \right]
=\frac{\sqrt{6}a i}{2(\sqrt\alpha-\sqrt{3\beta})\Mpl k}
{\cal Y} \left[1+{\cal O} \left( \frac{aH}{k} \right)\right]\,,
\label{eq:calZ_ini0}\\ 
{\cal B} &=& \frac{\sqrt{6}a^2}{4\Mpl k^2 \sqrt{\alpha}} 
\left( {\cal Y}+{\cal Z} \right)=
\frac{\sqrt{6}a^2}{2(\sqrt\alpha-\sqrt{3\beta})\Mpl k^2}\,
{\cal Y} \left[1+{\cal O} \left( \frac{aH}{k} \right)\right]
\,, \label{eq:calZ_ini}
\ea
where we used the relation (\ref{ZYre}). 
Recall that ${\cal Y}$ is given by Eq.~(\ref{calY}). 
The ratio between the leading-order terms to ${\cal A}$ and ${\cal B}$ 
is ${\cal B}/{\cal A}=-ia/k$, 
so the amplitude $|H{\cal B}/{\cal A}|$ is of order $aH/k \ll 1$ for sub-Hubble modes.
We will use Eqs.~(\ref{eq:calZ_ini0}) and (\ref{eq:calZ_ini}) as the initial conditions of perturbations 
for the modes deep inside the Hubble radius.

\subsection{Numerical integration}
\label{numesec}

In the flat gauge, we numerically integrate the closed fourth-order differential equations 
for ${\cal A}$ and ${\cal B}$ together with those 
of $\Psi$, $\Phi$, and $\zeta$. 
For this purpose, we introduce the following quantities:
\be
h = \frac{H}{H_i}\,,\qquad 
K = \frac{k}{a H_i}\,,\qquad
\bar{\alpha}=\alpha H_i^2\,,\qquad
\bar{\beta}=\beta H_i^2\,,\qquad
\chi=H_i {\cal B}\,,
\ee
where $H_i$ is the Hubble parameter at the onset of 
integration (with scale factor $a_i$).
We also define the following perturbed variables
\be
{\cal A}_k=\frac{k^{3/2}}{\sqrt{2\pi^2}} {\cal A}\,,
\qquad 
{\cal B}_k=\frac{k^{3/2}}{\sqrt{2\pi^2}} {\cal B}\,,
\qquad 
\chi_k=H_i {\cal B}_k\,,
\qquad
\Psi_k=\frac{k^{3/2}}{\sqrt{2\pi^2}} \Psi\,,
\qquad
\Phi_k=\frac{k^{3/2}}{\sqrt{2\pi^2}} \Phi\,,
\qquad
\zeta_k=\frac{k^{3/2}}{\sqrt{2\pi^2}} \zeta\,.
\ee
Then, the two gravitational potentials 
can be expressed as
\be
\Psi_k={\cal A}_k + h \chi_k'\,,
\qquad \mathrm{and} \qquad
\Phi_k=h \chi_k\,,
\label{Psire}
\ee
where a prime in this section denotes the differentiation 
with respect to the number of $e$-foldings $N=\ln (a/a_i)$. 
Together with solving the perturbation equations of motion, 
we integrate the following background equations of motion 
\be
h'=-h \epsilon\,,
\qquad
\epsilon' = \frac{3}{2} \epsilon^2 
- 3\epsilon +\frac{1}{12\bar{\beta} h^2}\,.
\ee
The initial value of $\epsilon$ is chosen to realize the sufficient 
number of $e$-foldings ($N>70$) during inflation. 
On using Eqs.~(\ref{eq:calZ_ini0}) and (\ref{eq:calZ_ini}), 
we choose the initial conditions of ${\cal A}_k$ and $\chi_k$ at $N=0$, 
as\footnote{All the perturbations labeled by $k$, 
for instance, $\chi_k$, $\zeta_k$, etc., satisfy the same closed 
fourth-order differential equation as their unnormalized counterparts, $\chi$, $\zeta$, etc., with different initial conditions only 
by the factor $k^{3/2}/\sqrt{2\pi^2}$.}
\ba
& &
{\cal A}_k(0)=iK_i 
\frac{|\bar\alpha-3\bar\beta|^{1/2}\,
{\rm sign}\bigl(\sqrt{\bar\alpha}-\sqrt{3\bar\beta}\bigr)}
{6\sqrt{2}\pi \sqrt{\bar\alpha\bar\beta}} 
\frac{H_i}{\Mpl}\,,
\label{Ak0}
\qquad
\frac{\rd^n {\cal A}_k}{\rd N^n}(0) 
\simeq (-iK_i)^n\,{\cal A}_k(0)\,,\label{eq:A_k}\\
& &
\chi_k(0)=\frac{|\bar\alpha-3\bar\beta|^{1/2}\,
{\rm sign}\bigl(\sqrt{\bar\alpha}-\sqrt{3\bar\beta}\bigr)}
{6\sqrt{2} \pi \sqrt{\bar\alpha\bar\beta}} 
\frac{H_i}{\Mpl}\,,
\qquad
\frac{\rd^n\chi_k}{\rd N^n}(0) \simeq (-iK_i)^n
\chi_k(0)\,.\label{eq:chi_k0}
\ea
where $K_i=K(0)=k/(a_iH_i)$. 
Notice that there is a simple relation 
${\cal A}_k (0)=iK_i\,\chi_k(0)$ for the leading-order 
solution.
The ratio $H_i/\Mpl$ and the couplings 
$\alpha$, $\beta$ determine the initial amplitude of 
$\chi_k$. The typical Hubble scale for 
Starobinsky inflation is  
$H_i/\Mpl={\cal O}(10^{-5})$, so that 
$|\chi_k(N=0)|={\cal O}(10^{-6})$ for 
the couplings $\alpha$ and $\beta$ whose 
orders are similar to each other. 
The initial conditions of gravitational potentials are
$\Psi_k(0)={\cal A}_k(0)+\chi_k'(0)$ 
and $\Phi_k(0)=\chi_k(0)$.

From Eq.~(\ref{eq:zeta_FG}), the curvature perturbation $\zeta_k$  
can be expressed, in terms of $\mathcal{A}_k$ and $\chi_k$, as
\be
\zeta_k=\frac{4 \bar\beta \left[h \left(2\chi_k +\chi_k'\right)
+\mathcal{A}_k \right]}
{1-6 \bar{\beta} \epsilon \left(\epsilon -2\right) h^{2}}K^2
-\frac{12\bar\beta h^{2} 
\left[2 \mathcal{A}_k\, (2 -\epsilon) +\mathcal{A}_k'\right] 
}{1-6  \bar{\beta} \epsilon\left(\epsilon -2\right) h^{2}}\,.
\label{zetare}
\ee
From Eq.~(\ref{Ak0}), we find that 
${\cal A}_k(0)$ is proportional to 
$K_i H_i/\Mpl$ and hence $\zeta_k(0)$ contains a large term $K_i^3 (H_i/\Mpl)$ 
for the sub-Hubble modes $K_i \gg 1$. 
Because of the relation ${\cal A}_k (0)=iK_i\,\chi_k(0)$, this term 
exactly cancels the other contribution $\chi_k'(0) K_i^2$ for the 
leading-order initial condition $\chi_k'(0)=-iK_i \chi_k(0)$. 
This cancellation implies that we need to take into account 
terms of order ${\cal O}(a^2 H^2/k^2)$ in Eq.~(\ref{ZYre}) to 
estimate the initial value of $\zeta_k(0)$ correctly.
We also note that Eq.~(\ref{zetare}) contains the $N$ 
derivatives of ${\cal A}_k$ and $\chi_k$. Since 
the amplitudes of ${\cal A}_k$ and $\chi_k$ change in time, 
we take the $N$ derivatives of these fields 
without neglecting their time dependence. 
These precise manipulations show that the leading-order contributions 
to the first and second terms in Eq.~(\ref{zetare}) cancel each other with 
respect to the large $K_i$ expansion. 
Then, the initial value of $\zeta_k(0)$ is typically 
of order $\bar{\beta}\chi_k (0)K_i$. 
After deriving the precise numerical value of $\zeta_k(0)$ from Eq.~(\ref{zetare}) 
without using the approximation, the $N$ derivatives of 
$\zeta_k$ at $N=0$ can be estimated 
as $(\rd^n\zeta_k/\rd N^n)(0) \simeq (-iK_i)^n\,\zeta(0)$.

\begin{figure}[ht]
\centering
\includegraphics[width=0.8\textwidth]{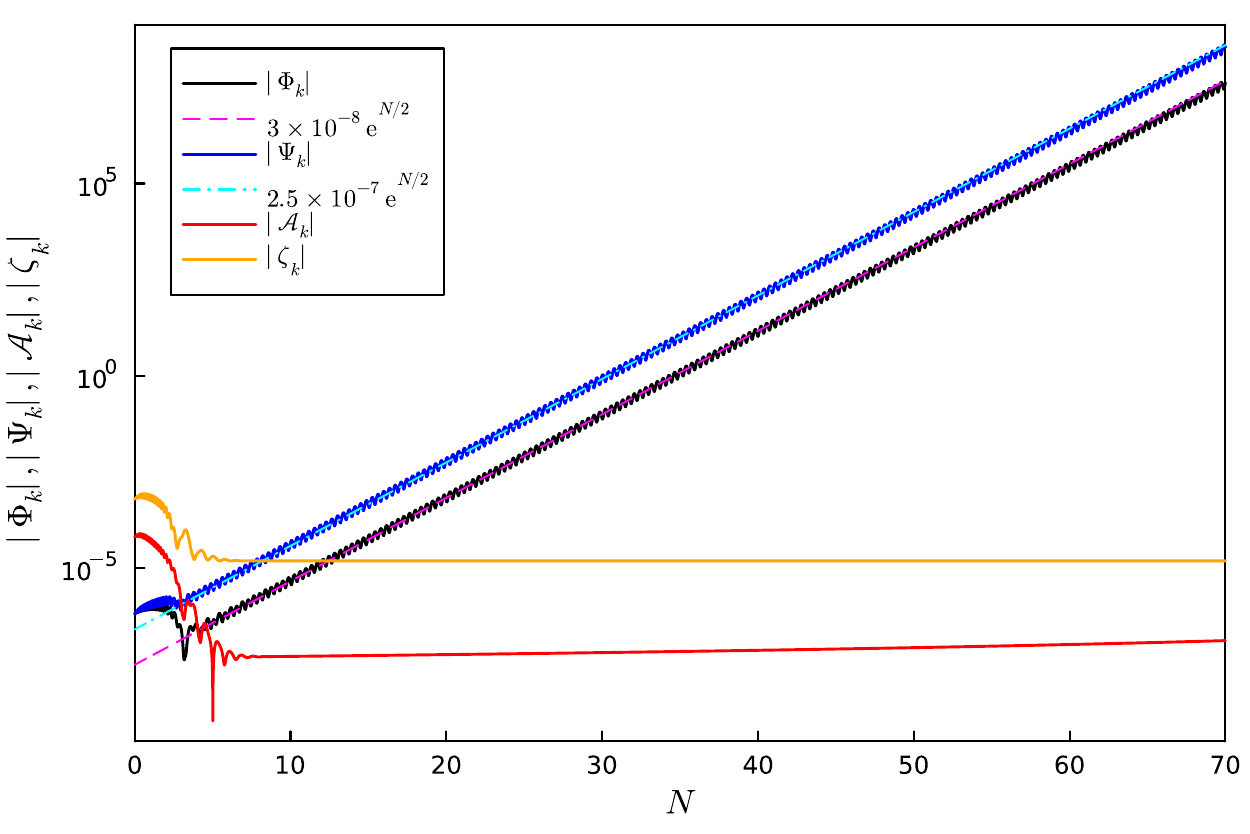}
\caption{Exponential increase of $|\Phi_k|=|h \chi_k|$ and $|\Psi_k|$ 
for the couplings $\bar{\alpha}=1$ and $\bar{\beta}=6$. 
In the regime $k \ll aH$, the growth of both $|\Phi_k|$ and $|\Psi_k|$ 
can be well-fitted by an exponential function proportional to $e^{N/2}$. 
The initial conditions are chosen to be $h(0)=1$, 
$\epsilon(0)=0.0047$, $H_i/\Mpl=10^{-5}$, and 
$\Phi_k(0)=\chi_k(0) \simeq -6.31\times10^{-7}$ [which is 
determined by Eq.~(\ref{eq:chi_k0})] for the sub-Hubble mode $K(0)=100$.
We also show two exponential functions proportional to $e^{N/2}$ 
as a dashed pink line and a dash-dotted 
light blue line, 
which fit well with the numerical solutions of $|\Phi_k|$ and $|\Psi_k|$, respectively. 
The initial conditions of ${\cal A}_k$ and $\zeta_k$ are known from Eqs.~(\ref{Ak0}) and (\ref{zetare}), as
${\cal A}_k(0)\simeq -6.34\times10^{-7}-6.32\times10^{-5}i$ 
and $\zeta_k(0) \simeq 1.3\times10^{-4}-5.9\times 10^{-4}i$. 
To show the evolution of $\zeta_k$, we have solved its own closed 
fourth-order differential equation. 
We see that $|\zeta_k|$ approaches 
a constant after the Hubble radius crossing, 
while ${\cal A}_k$ exhibits very 
mild growth.}
\label{fig:all}
\end{figure}

As we will show in the numerical calculation below, the perturbation 
$\chi_k$ is unstable, but the 
decrease of the $K^2$ term (proportional to $a^{-2}$) 
in Eq.~(\ref{zetare}) suppresses the growth of $\chi_k K^2$.
Therefore, after the Hubble radius crossing, 
$\zeta_k$ depends mostly on $\mathcal{A}_k$ and 
its $N$-derivative. Hence, if the growth of ${\cal A}_k$ 
is insignificant, this is also the case for $\zeta_k$. 
Furthermore, this shows that, in the super-Hubble regime, 
$\zeta_k$ is related only to ${\cal A}_k$ and vice versa. 
Thus, the whole scalar sector, which consists of two independent 
dynamical d.o.f.s, cannot be described by 
$\zeta_k$ and ${\cal A}_k$ alone in the regime 
$k/a \ll H$.

In Fig.~\ref{fig:all}, we plot the evolution of $|\Phi_k|=|h \chi_k|$, $|\Psi_k|$, $|\mathcal{A}_k|$, and $|\zeta_k|$ for $\bar{\alpha}=1$, 
$\bar{\beta}=6$, and $K(0)=100$ with 
the slow-roll parameter $\epsilon(0)=0.0047$. The initial conditions for $\chi_k$ are instead chosen as to fulfill Eq.\ \eqref{eq:chi_k0}, with $H_i=10^{-5}\Mpl$.
We solved the fourth-order differential equations 
of $\chi_k$ and ${\cal A}_k$ and computed 
$\Phi_k$, $\Psi_k$, and $\zeta_k$ by exploiting the 
relations (\ref{Psire}) and (\ref{zetare}). 
We also performed the direct integration of the fourth-order differential equations 
for $\Phi_k$, $\Psi_k$, and $\zeta_k$ by implementing the initial conditions for each of them in terms of those given for $\chi_k(0)$ and found that the results are in perfect agreement with those computed from $\chi_k$ and ${\cal A}_k$.

In Fig.~\ref{fig:all}, we observe that both the amplitudes of $\Phi_k=H{\cal B}_k$ and $\Psi_k$ grow in proportion to $e^{Ht/2} \simeq e^{N/2}$. 
We recall that we used the approximation $\beta H^2 \gg 1$ to derive the analytic solutions (\ref{Phiso1}), but the numerical results show that this estimation is valid even for $\beta H^2$ of order 1. 
In Fig.~\ref{fig:all}, the gravitational potentials exceed the order 1 
around the $e$-foldings $N=30 \sim 35$ after the onset of inflation. 
Thus, the exponential growth of ${\cal B}$ in the perturbed line 
element (\ref{metflat}) invalidates the FLRW background. 
Due to the uncertainty principle, the initial value of the perturbation 
$\chi_k$ has a nonvanishing value related to the energy scale 
$H_i/\Mpl$ during inflation. Since $H_i/\Mpl$ should not be 
much smaller than $10^{-5}$, the gravitational potentials reach order 1 after the amplification of $e^{N/2}$ times with $N>30$.
As we estimated analytically in Sec.~\ref{scasec}, the perturbation
$\zeta_k$ approaches a constant after the Hubble radius crossing, 
while ${\cal A}_k$ shows very mild growth.
We recall that $\zeta_k$ and ${\cal A}_k$ are related to each other in the super-Hubble regime and that they are not sufficient to describe the dynamics of scalar perturbations with two propagating d.o.f.s. 
Indeed, we cannot eliminate the instability of the other dynamical perturbation ${\cal B}_k=\Phi_k/H$.

\begin{figure}[ht]
\centering
\includegraphics[width=0.8\textwidth]{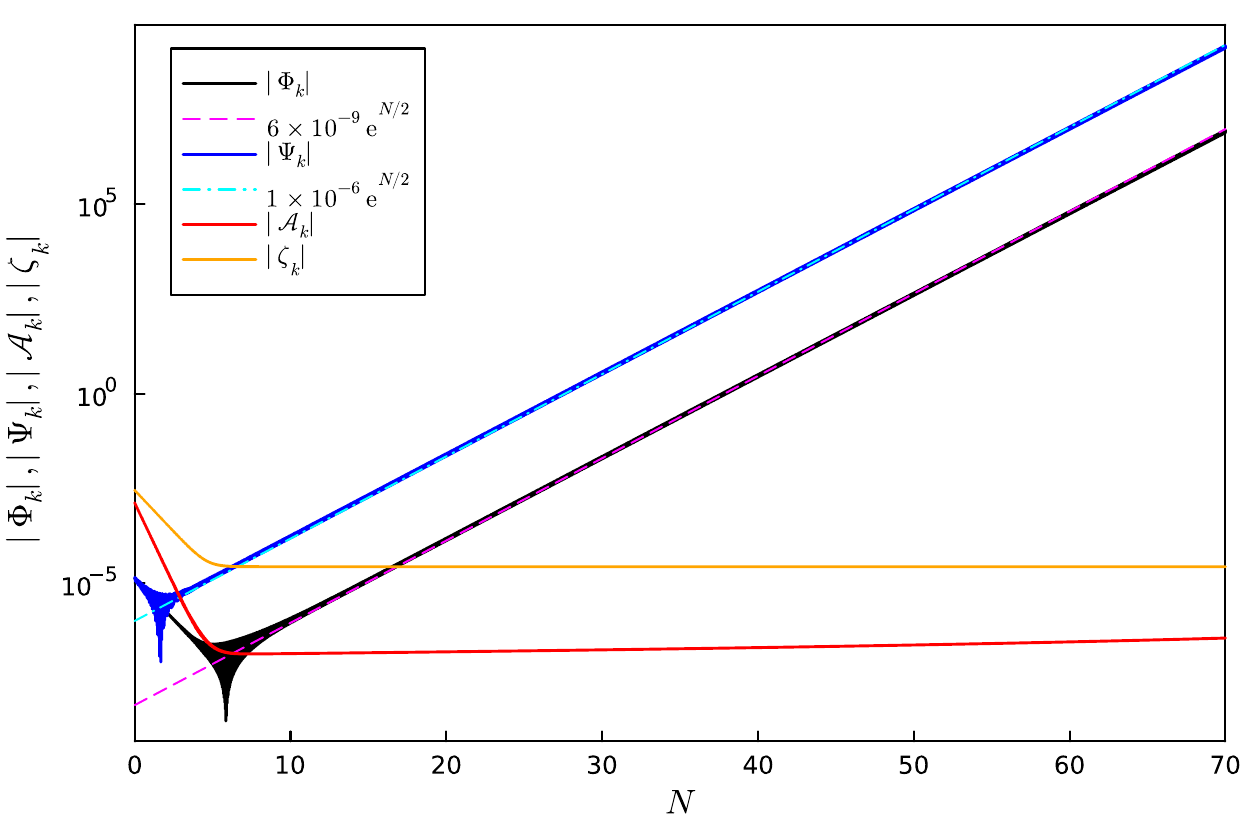}
\caption{The same as Fig.~\ref{fig:all}, 
but for the Weyl coupling $\bar\alpha=1/400$, 
$\chi_k(0) \simeq -1.3\times10^{-5}$, $\mathcal{A}_k(0) \simeq -1.3\times10^{-5} - 0.0013 i$, 
and $\zeta_k(0)\simeq 0.0028 - 2.86\times10^{-5}i$. 
Also for this small coupling constant $\bar\alpha$, 
the gravitational potentials 
grow as $e^{N/2}$.}
\label{fig:allas}
\end{figure}

\begin{figure}[ht]
\centering
\includegraphics[width=0.8\textwidth]{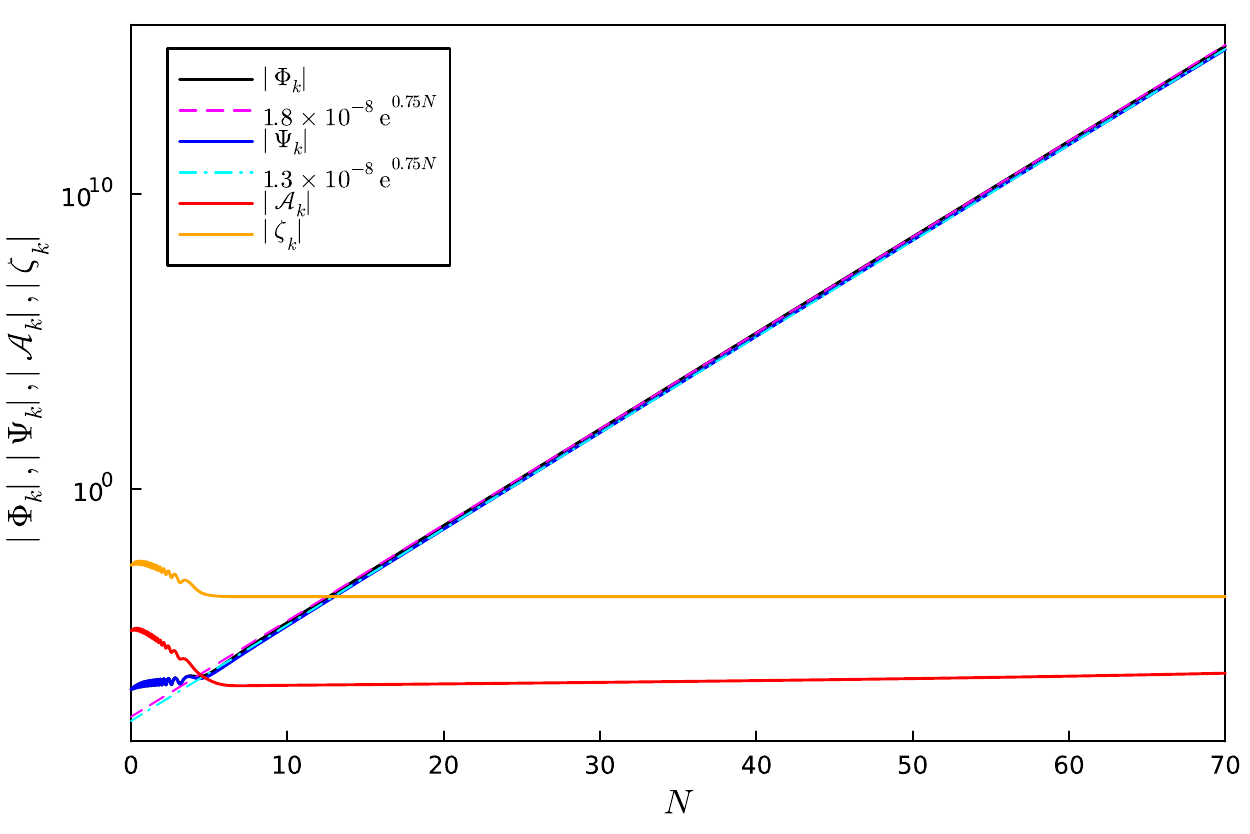}
\caption{The same as Fig.~\ref{fig:all}, 
except for $\bar{\alpha}=400$, 
$\chi_k(0) \simeq 1.5\times 10^{-7}$, $\mathcal{A}_k(0) \simeq 1.5\times10^{-7} 
+ 1.5\times10^{-5}i$, 
$\zeta_k(0)\simeq -3.2\times10^{-5} - 0.0025i$, and for the Weyl coupling $\bar{\alpha}=400$. 
For this large coupling constant $\bar{\alpha}$, 
the gravitational potentials grow as fast as $e^{3N/4}$.}
\label{fig:allal}
\end{figure}

In Fig.\ \ref{fig:allas}, we show the evolution of 
gauge-invariant perturbations by keeping the same initial conditions and model parameters as those in Fig.~\ref{fig:all}, except for $\bar\alpha$ which is set to a smaller value ($\bar{\alpha}=1/400$) 
and for the initial values of the fields $\chi_k$, ${\cal A}_k$, and $\zeta_k$. Even with this small value of the Weyl coupling, 
both $|\Phi_k|$ and $|\Psi_k|$ increase in proportion to 
$e^{Ht/2} \simeq e^{N/2}$ after the Hubble radius 
crossing, while the growth of $|\zeta_k|$ and $|{\cal A}|$ 
is suppressed. Indeed, this behavior is expected according to the analytic estimations of $\Phi_k=H{\cal B}_k$, $\Psi_k$, ${\cal A}_k$, and $\zeta_k$. 
For small values of $\alpha$, the last two terms in Eqs.~(\ref{Psiso}) and (\ref{Phiso}) grow in proportion to $e^{Ht/2}$ with oscillations.\footnote{In Fig.~\ref{fig:allas}, the oscillations 
of $|\Psi_k|$ and $|\Phi_k|$ are not clearly seen at large $N$, but we confirmed that they are present by enlarging the figure.} 
Note that Starobinsky inflation without the Weyl term ($\alpha=0$) cannot be recovered by simply taking the limit $\alpha \to 0$ in our theory.
For $\alpha=0$ there is only a single scalar d.o.f. arising from 
the $\beta R^2$ term, in which case the exponential growth 
of gravitational potentials is absent. 
However, as we saw for the vector and tensor perturbations, the squared 
mass of the extra modes, when the Weyl-squared term is present, is typically of order $\alpha^{-1}$. Then, in the limit $\alpha \to 0$, the extra modes become very massive and we should expect them to acquire a mass larger than an ultraviolet cutoff scale of order $\Mpl$.
In this case, they may be integrated out from the theory. 
On the other hand, the instability of scalar perturbations persists for the Weyl coupling constant in the range $\alpha \gtrsim \Mpl^{-2}$. 

In Fig.~\ref{fig:allal}, the evolution of gauge-invariant perturbations is plotted for the large Weyl coupling $\bar{\alpha}=400$ with $\bar{\beta}=6$. 
In this case, the condition $\alpha>48 \beta$ is satisfied and hence the two gravitational potentials grow 
as Eq.~(\ref{Phiso2}) without oscillations.
Indeed, our numerical results in Fig.~\ref{fig:allal} demonstrate that, after the Hubble radius crossing, the growth of $|\Phi_k|$ and $|\Psi_k|$ occurs faster than in the case $\bar{\alpha}=1$. 
The suppressed growth of the other perturbations ${\cal A}_k$ 
and $\zeta_k$ also agrees with the analytic estimation 
in the regime $\beta H^2 \gg 1$.

In general, out of two exponentially growing modes 
$\Phi_k$ and $\Psi_k$, one can find some linear combinations of them, 
like  $\zeta_k$, whose growth is suppressed in the super-Hubble regime. 
In this theory, however, two scalar perturbations determine the stability of the background and not only one. 
Therefore, the stability of one linear combination is not sufficient for guaranteeing the stability of 
the whole dynamical system. In the flat gauge, the instability of the field ${\cal B}$ appearing 
in the perturbed metric (\ref{metflat}) is enough to make the whole background unstable 
in the super-Hubble regime. 
We also note that, in the flat gauge, the Weyl tensor component ${C^{0}}_{i0j}$ in real space 
can be expressed as 
\be
{C^{0}}_{i0j}=-\frac{1}{2} \left( \partial_{i} \partial_{j}-\frac{1}{3} \delta_{ij} \nabla^2 
\right) ({\cal A}+\dot{\cal B}-H{\cal B})\,,
\ee
which vanishes on the FLRW background.  
In Fourier space, we can also consider the evolution of the perturbation 
$C_k \equiv {\cal A}_k+\dot{\cal B}_k-H{\cal B}_k$ to see the departure from the background. 
While ${\cal A}_k$ does not exhibit the exponential growth, the amplitude of 
${\cal B}_k=\Phi_k/H$ evolves as ${\cal B}_k=b_0\,e^{\lambda H t}$ 
after the Hubble radius crossing, where $b_0$ and $\lambda$ are nonzero 
constants and $H$ is assumed to be constant during inflation.
{}From Eqs.~(\ref{Phiso1}) and (\ref{Phiso2}), the power $\lambda$ is 
in the range $1/2 \le \lambda <1$. 
Since the amplitude of $\dot{\cal B}_k-H{\cal B}_k$ has the time dependence 
$\dot{\cal B}_k-H{\cal B}_k=b_0 (\lambda-1)H e^{\lambda Ht} \neq 0$ for super-Hubble 
modes, the perturbation $C_k$ grows exponentially to spoil the FLRW background. 
Indeed, we have numerically found the exponential growth of the amplitude of $C_k$.
 
We also solved the perturbation equations in the Newtonian gauge and obtained the same 
numerical solutions for the gauge-invariant fields $\Psi_k$, 
$\Phi_k$, $\zeta_k$, and ${\cal A}_k$ as those in the flat gauge.
In the Newtonian gauge, the violation of the FLRW background occurs by the growth 
of two gravitational potentials $\Psi$ and $\Phi$ in the perturbed line element (\ref{Newtonian}). 
For this gauge choice, the Weyl tensor component ${C^{0}}_{i0j}$ in real space 
is given by \cite{Durrer:2020fza}
\be
{C^{0}}_{i0j}=-\frac{1}{2} \left( \partial_{i} \partial_{j}-\frac{1}{3} \delta_{ij} \nabla^2 
\right) (\Psi-\Phi)\,.
\ee
As we can see in Figs.~\ref{fig:all} and \ref{fig:allas}, the gravitational potentials 
$\Psi_k$ and $\Phi_k$ in Fourier space are generally different from each other. 
Hence, the amplitude of the combination $\Psi_k-\Phi_k$ also grows exponentially 
after the Hubble radius crossing. 
For increasing $\alpha$ relative to $\beta$, the difference between $\Psi_k$
and $\Phi_k$ tends to be smaller (see Fig.~\ref{fig:allal}). 
Provided that $\alpha$ is finite, however, the exponential growth of ${C^{0}}_{i0j}$ 
always occurs by reflecting the fact that the amplitude of ${C^{0}}_{i0j}$ 
is proportional to $|\lambda-1| H e^{\lambda Ht} $ with $1/2 \le \lambda<1$,
as we discussed for the flat gauge above.
Indeed, irrespective of the values of $\alpha$, 
we numerically confirmed the exponential increase of $|\Psi_k-\Phi_k|$ 
in the Newtonian gauge.

In the unitary gauge, we numerically observed the same exponential growth 
of $\Phi_k$ and the constancy of $\zeta_k$ after the Hubble radius crossing.
In this case, the gauge-invariant combination 
${\cal B}_u=(\Phi-\zeta)/H$ appearing in the perturbed 
metric (\ref{unimet}) is subject to the exponential growth, 
thereby invalidating the FLRW background.

\section{Conclusions}
\label{consec}

In this paper, we studied the dynamics of cosmological perturbations during inflation in quadratic gravity containing the Weyl term $-\alpha C^2$ besides the Ricci squared term $\beta R^2$ in the action. Although the Weyl curvature does not affect the background inflationary dynamics driven by the $\beta R^2$ term, the evolution of perturbations is modified by the presence of derivatives higher than second order. 
Since these higher-order derivatives can give rise to ghosts, it is of interest to explore whether or not the ghosts can lead to instabilities of the FLRW background. 

As we discussed in Sec.~\ref{backsec}, geometric inflation is 
realized by the $\beta R^2$ term with $\beta>0$, 
where the coupling constant $\beta$ is related to the mass 
squared $m_S^2$ of a new scalar d.o.f. (scalaron) as $\beta=1/(6m_S^2)$. 
To realize the number of $e$-foldings larger than 60, we require 
that the Hubble parameter $H_i$ at the onset of inflation is 
in the range $\beta H_i^2 \gtrsim {\cal O}(1)$.
If we transform the action (\ref{Saction}) to that in the Einstein 
frame, the quadratic gravity can be interpreted as the conformally 
invariant Weyl theory in the presence of a canonical scalaron field
with the potential. Unlike the past related works \cite{Clunan:2009er, Deruelle:2010kf, Myung:2015vya, Ivanov:2016hcm, Anselmi:2021rye}, we have carried out all the analysis in the physical Jordan frame.

In Sec.~\ref{vecsec}, we showed that the Weyl term gives rise to two dynamical vector d.o.f.s propagating with the speed of light. 
For the Weyl coupling $\alpha>0$ the two ghosts are present with the positive mass squared $m_W^2=1/(2\alpha)$, while, for $\alpha<0$, there are no ghosts. 
In the latter case, however, the negative value of $m_W^2$ 
leads to the tachyonic instability of vector perturbations 
for $|\alpha|$ at most of order $\beta$. 
To avoid such an instability which violates the inflationary FLRW background, 
we demand the condition $\alpha>0$ at the expense of admitting the existence of ghosts.

In Sec.~\ref{tensec}, we derived the second-order action of 
tensor perturbations and introduced Lagrange multiplier fields 
$\chi_i$ ($i=1,2$) associated with higher-order time derivatives.
There are four dynamical d.o.f.s in the tensor sector, two of which 
behave as ghosts. 
Using the WKB approximation for the modes 
deep inside the Hubble radius ($k/a \gg H$), the speed of 
tensor perturbations is equivalent to 1 with vanishing masses.
Despite the presence of the Weyl ghost, 
the classical perturbations are not subject to either Laplacian 
or tachyonic instabilities for subhorizon modes.
In the super-Hubble regime  ($k/a \ll H$), 
tensor perturbations $h_i$ obey the fourth-order differential 
Eq.~(\ref{eq:GWs}). Provided that the couplings $\alpha$ 
and $\beta$ are in the ranges $\alpha>0$ 
and $\beta H^2 \gtrsim 1$, we showed that $h_i$'s 
approach constants after the Hubble radius crossing. 
This means that, despite the presence of ghosts, 
tensor perturbations are subject to neither Laplacian nor 
tachyonic instabilities.

In Sec.~\ref{scasec}, we studied the stability and evolution of 
scalar perturbations by choosing several different gauge conditions. 
There are two dynamical propagating d.o.f.s in the scalar sector 
arising from the Lagrangians $-\alpha C^2$ and $\beta R^2$.
For $\alpha>0$ and $\beta>0$, the scalaron is not a ghost, 
but the other dynamical mode behaves as a ghost. 
To study the dynamics of perturbations, we also introduced 
several gauge-invariant perturbations such as those defined in Eqs.~(\ref{gravi})--(\ref{zeta}).
We chose the Newtonian, flat, and unitary gauges and 
derived the closed differential equations for $\Psi$, $\Phi$, 
${\cal A}$, ${\cal B}=\Phi/H$, and $\zeta$. 
We found that the coefficients of these differential equations are uniquely fixed independent of the gauge choices. 
We showed that, after the Hubble 
radius crossing, both $\Psi$ and $\Phi$ grow 
exponentially, while ${\cal A}$ and $\zeta$ approach 
constants. 

In the Newtonian gauge given by the perturbed 
line element (\ref{Newtonian}), the exponential 
growth of $\Psi$ and $\Phi$ occurs in the $g_{00}$ 
and $g_{ii}$ metric components. 
This violates the stability of the FLRW background 
after the perturbations cross the Hubble radius during inflation. 
For the flat-gauge line element (\ref{metflat}) the growth of ${\cal A}$ is suppressed, but the exponential increase of ${\cal B}=\Phi/H$ occurs together with the enhancement of $\Phi$. 
We have also numerically confirmed this behavior for the gauge-invariant perturbations in the numerical simulations of Figs.~\ref{fig:all}--\ref{fig:allal} performed in Sec.~\ref{sec:numerics}.
In the unitary gauge, the perturbed line element (\ref{unimet}) 
also contains the instability mode $\Phi$ in the $g_{0i}$ 
metric component. 
We stress that these instabilities are the physical ones 
arising from the gravitational interaction between 
the scalaron and the other ghost d.o.f.

We have thus shown that the inflationary FLRW background realized by the $\beta R^2$ term is violated by the presence of the Weyl term. In other words, the Universe becomes highly inhomogeneous during inflation, being 
incompatible with the observations of CMB temperature anisotropies. This instability of scalar perturbations 
is present for the wide coupling range 
$\alpha \gtrsim \Mpl^{-2}$ in which the mass term 
$1/\sqrt{\alpha}$ associated with the Weyl term 
does not exceed the ultraviolet scale of order $\Mpl$. 
Unless the scalar ghost arising from the Weyl term is suitably eliminated as a physical propagating d.o.f. and the classical instability of the background is removed, the quadratic curvature theory with $\alpha \neq 0$ is excluded as a viable model of inflation (or at most, the coupling $\alpha$ must be so small that the mass of the extra modes becomes larger than the cutoff of the theory).
Related to the ghost issue, there is an approach of ``fakeon'' where the ghost does not appear as a physical state after quantizing it as a fake 
d.o.f. \cite{Anselmi:2017ygm, Anselmi:2018kgz, Anselmi:2020lpp}.
There are also some approaches to the ghost problem in quantum field theory by keeping its physical status intact \cite{Bender:2007wu, Donoghue:2019fcb, Salvio:2020axm, Kubo:2022dlx}. 
In such approaches to the ghost problem, it will be of interest to study the stability of cosmological perturbations and resulting observational consequences in detail.

\section*{Acknowledgments}

The work of A.D.F. was supported by the Japan Society for the Promotion of Science Grants-in-Aid for Scientific Research No.\ 20K03969 and by Grant No. PID2020-118159GB-C41 funded by MCIN/AEI/10.13039/501100011033. 
S.T. was supported by the Grant-in-Aid for Scientific Research 
Fund of the JSPS No.~22K03642 and Waseda University 
Special Research Project No.~2023C-473.

\bibliographystyle{mybibstyle}
\bibliography{bib}

\end{document}